\documentclass[fleqn,usenatbib]{rasti}

\usepackage{newtxtext,newtxmath}

\usepackage[T1]{fontenc}

\DeclareRobustCommand{\VAN}[3]{#2}
\let\VANthebibliography\thebibliography
\def\thebibliography{\DeclareRobustCommand{\VAN}[3]{##3}\VANthebibliography}

\usepackage{graphicx}
\usepackage{amsmath}
\usepackage{booktabs}
\usepackage{hyperref}
\usepackage{float}
\usepackage{dblfloatfix}
\usepackage{subcaption}
\usepackage{xcolor}
\usepackage{placeins}  
\usepackage[normalem]{ulem}
\usepackage{bm}
\usepackage{algorithm}
\usepackage{algpseudocode}
\usepackage{array}
\usepackage{listings}
\usepackage{tikz}
\usetikzlibrary{arrows.meta, positioning, fit, backgrounds, calc, shapes.geometric}
\title[JAX-Accelerated Parametric CMB Component Separation]{A GPU-Accelerated JAX Framework for Robust Parametric Component Separation and Clustering Optimization for CMB Polarization Satellites}
\author[W. Kabalan et al.]{Wassim Kabalan$^{1}$\thanks{Contact: \href{mailto:wassim@apc.in2p3.fr}{wassim@apc.in2p3.fr}}, 
Arianna Rizzieri$^{2}$,
Wuhyun Sohn$^{1}$,
Artem Basyrov$^{1}$,
Alexandre Boucaud$^{1}$,
\newauthor
Benjamin Beringue$^{1}$,
Pierre Chanial$^{1}$,
Ema Tsang King Sang$^{1}$,
Josquin Errard$^{1}$\thanks{Contact: \href{mailto:josquin@apc.in2p3.fr}{josquin@apc.in2p3.fr}}
\\
$^{1}$Université Paris Cité, CNRS, Astroparticule et Cosmologie, F-75013 Paris, France \\
$^{2}$Department of Physics, University of Oxford, Denys Wilkinson Building, Keble Road, Oxford OX1 3RH, United Kingdom
}

\date{Accepted XXX. Received YYY; in original form ZZZ}
\pubyear{\the\year{}}

\begin{document}
\label{firstpage}
\pagerange{\pageref{firstpage}--\pageref{lastpage}}
\maketitle

\begin{abstract}
We present a novel, \texttt{JAX}-powered implementation of a parametric component-separation method for CMB polarization data, explicitly designed to handle spatially varying foreground Spectral Energy Distributions (SEDs). The approach models this variation across the sky by grouping sets of pixels that share common foreground spectral parameters, scanning over thousands of such configurations to evaluate the trade-off between model complexity and residual systematic contamination. Built within the \texttt{FURAX} framework which is a \texttt{JAX}-powered environment for CMB data analysis. Our pipeline extends the \texttt{fgbuster} parametric formalism.
It enables fully vectorized, GPU-accelerated evaluation of the spectral likelihood, map reconstruction, and diagnostic metrics across tens of thousands of pixel subset configurations, noise realizations, and sky regions. Our implementation achieves up to ${\sim}100\times$ speed-up over the \texttt{scipy} TNC optimizer used in \texttt{fgbuster} when running on GPUs, as well as giving more robust results. When applied to LiteBIRD-like simulations with spatially varying foreground SEDs, our optimized K-means configuration reduces the 68\% upper limit on the tensor-to-scalar ratio $r$ by $\approx 30\%$ relative to a fixed, previously derived multi-resolution configuration, while maintaining competitive statistical uncertainties.
\end{abstract}

\begin{keywords}
cosmic background radiation -- methods: data analysis -- methods: statistical -- gravitational waves -- inflation
\end{keywords}



\section{Introduction}

Detecting the B-mode polarization in the Cosmic Microwave Background (CMB) is a central objective in modern cosmology, as it could offer direct observational evidence of primordial gravitational waves generated during the Universe’s inflationary period~\citep{Kamionkowski1997, Seljak1997}. The amplitude and shape of this B-mode signal are direct probes of inflationary energy scales, potentially reaching physics at $\sim10^{16}$\,GeV. Upcoming experiments, such as LiteBIRD~\citep{LiteBIRD2022}, aim to measure these faint signals with unprecedented sensitivity, targeting constraints on the tensor-to-scalar ratio $r$ down to levels of $< 0.001$. Achieving these ambitious goals is complicated by the presence of polarized astrophysical foregrounds, primarily arising from Galactic synchrotron emission and Galactic thermal dust emission~\citep{Planck2016SED, MeisnerFinkbeiner2015}. These foregrounds dominate the CMB polarization signal across much of the sky and exhibit spatially varying spectral properties~\citep{Planck2018IV}. Large angular scales (low multipoles) are both most sensitive to primordial B modes and most affected by foregrounds and instrument systematics~\citep{LiteBIRD2022}, making accurate separation of the CMB signal from these other components especially critical in this regime.

Component separation methods are therefore a necessary data analysis step for experiments targeting primordial B modes. They can be broadly classified as parametric (assuming a model for the spectral energy distributions (SEDs) of foregrounds parameterized by the so-called ``spectral parameters''~\citep{Brandt1994, Eriksen2006, Stompor_2009}) or non-parametric, which rely on statistical independence or internal templates~\citep{Delabrouille2003, Cardoso2008}. Key tools include \textsc{Commander}~\citep{Eriksen2008, Planck2018IV}, \textsc{SMICA}~\citep{Cardoso2008, Planck2018IV}, \textsc{MICMAC}~\citep{morshed2024pixel}, \textsc{NILC}~\citep{Zhang_2024} and \textsc{FGBuster}~\citep{poletti2023fgbuster, Rizzieri_2025}, each providing different approaches to the trade-off between flexibility and tractability.

We consider the parametric class, whose intrinsic difficulty lies in defining suitable parametric scaling laws for the polarized foreground components. These are today known to be more complex than a single uniform modified black body for the polarized thermal dust emission and a single uniform power law for the synchrotron emission. Recent observations indicate that the spectral energy distributions of foregrounds can be considered either to vary across the sky~\citep{Planck2018IV, fuskeland2014spatial} or as a mixing of individually uniform components~\citep{gjerlow2026cosmoglobe}. We focus here on the former interpretation, developing a component separation procedure that addresses the spatial variability of foreground SEDs in a systematic, data-driven manner. Such variability can indeed induce significant bias on the estimated value of $r$ if under-parameterized (averaging over too large regions) or inflate statistical variance if over-parameterized (pixel-by-pixel fitting). Finding the precise bias-variance trade-off in this high-dimensional configuration space is a key challenge for next-generation B-mode experiments.

To account for this spatial variability, parametric models must fit for each spectral parameter across independent sky pixel subsets.
A promising way to do this, within the \texttt{fgbuster} framework, as used in~\cite{LITEBIRD_PTEP_2022} and \cite{Rizzieri_2025}, defines these pixel subsets as a number of spatially connected sky regions (or patches) given by \textsc{HEALPix}~\citep{Gorski2005} super-pixels. The specific number of patches for each spectral parameter is determined by optimizing over different patch configurations. Traditionally, exploring the vast space of possible sky segmentations has been computationally intractable, as optimizing for a single configuration in this large space would require tens of thousands of CPU hours. Previous works have hence either built a single configuration by relying on external data or by combining spatial and spectral information, or restricted the analysis to a handful of heuristic patch choices (e.g., predefined Galactic masks or geometric super-pixels)~\citep{grumitt2020hierarchical, Puglisi_2022, Rizzieri_2025}. Our goal is instead to broadly explore the space of possible sky segmentations, allowing for a fully data-driven discovery of the optimal bias-variance trade-off. To achieve this, a paradigm shift toward accelerated, differentiable computing is required.

The method we develop here builds on the \textsc{FGBuster} framework. An optimized implementation allows us to efficiently explore a large space of pixel subset configurations. Our component separation pipeline is implemented within \texttt{FURAX}~\citep{FURAX}, a \texttt{JAX}-powered framework for CMB data analysis, and can be viewed as a \texttt{JAX}-native generalization of the \textsc{FGBuster} parametric approach, built for end-to-end differentiability, GPU acceleration, and large-scale model selection. 

\texttt{FURAX} uses the high-performance computing capabilities of \texttt{JAX}~\citep{JAX} to enable scalable and reproducible component separation pipelines suited for next-generation satellite missions. We demonstrate this efficiency by exploring region configurations more flexible than \texttt{HEALPix} pixels. These regions are defined via a spherical K-means clustering algorithm~\citep{Lloyd1982,DhillonModha2001} implemented using the \texttt{jax-healpy} package~\citep{JAXHEALPY}, henceforth referred to as `patches'. This flexibility enables exploration of a large number of configurations and selection of a final model by minimizing the error on the measured $r$ value.

This paper is organized as follows. In Section~\ref{sec:methodology}, we outline the theoretical framework, detailing the parametric component separation model, the spatial clustering strategy, grid search selection metrics, and the estimation of the tensor-to-scalar ratio $r$. In Section~\ref{sec:implementation}, we describe the computational infrastructure, introducing the distributed \texttt{FURAX} framework and the high-performance \texttt{AdaTopK} optimizer used to maximize the spectral likelihood. In Section~\ref{sec:results}, we present our validation on synthetic data and demonstrate the framework's advantages in balancing bias and variance compared to existing techniques. Finally, in Section~\ref{sec:conclusion}, we summarize our findings and discuss implications for future observational missions.

\section{Methodology}
\label{sec:methodology}

\subsection{Parametric Component Separation}
The algebraic formulation of our implementation of the core component separation routines follows the standard parametric formalism of \citet{Stompor_2009} as implemented in \textsc{FGBuster}~\citep{poletti2023fgbuster, Rizzieri_2025}.

We model multi-frequency sky observations in each pixel $p$ as a linear combination of astrophysical components with additive (Gaussian) noise:
\begin{equation}
    \mathbf{d}_p = \mathbf{A}_p(\boldsymbol{\beta}_p)\,\mathbf{s}_p + \mathbf{n}_p,
    \label{eq:data_model}
\end{equation}
where:

\begin{itemize}
    \item \( \mathbf{d}_p \in \mathbb{R}^{N_d} \): observed data vector in pixel $p$, with \(N_d~\equiv~N_{\mathrm{frequencies}} \times N_{\mathrm{Stokes}}\);
    \item \( \mathbf{s}_p \in \mathbb{R}^{N_s} \): sky components at a reference frequency \( \nu_0 \), with \(N_s~\equiv~N_{\mathrm{components}} \times N_{\mathrm{Stokes}}\);
    \item \( \mathbf{A}_p(\boldsymbol{\beta}_p) \in \mathbb{R}^{N_d \times N_s} \): mixing matrix encoding spectral dependencies;
    \item \( \mathbf{n}_p \sim \mathcal{N}(0, \mathbf{N}_p) \in \mathbb{R}^{N_d} \): Gaussian noise with known covariance \( \mathbf{N}_p \).
\end{itemize}

Each column of the mixing matrix \( \mathbf{A}(\boldsymbol{\beta}) \) models the SED of a sky component. In practice, this includes a blackbody for the CMB, a modified blackbody (MBB) for thermal dust emission, and a power law for synchrotron radiation. The vector \( \boldsymbol{\beta} \) denotes the set of spectral parameters for the latter two components, \( \boldsymbol{\beta} = (\beta_{\rm d}, T_{\rm d}, \beta_{\rm s}) \).

Working in thermodynamic CMB units, the CMB SED is frequency independent so that \(A_{\rm cmb}(\nu) = 1\). For a channel with central frequency \(\nu\), the dust and synchrotron columns of the mixing matrix are written as
\begin{align}
    A_{\rm d}(\nu;\beta_{\rm d},T_{\rm d})
    &= \frac{\nu}{\exp\!\left(\frac{h\nu}{kT_{\rm d}}\right)-1}
       \frac{\exp\!\left(\frac{h\nu_0}{kT_{\rm d}}\right)-1}{\nu_0}
       \left(\frac{\nu}{\nu_0}\right)^{\beta_{\rm d}}, \label{eq:mixing_columns} \\ 
    A_{\rm s}(\nu;\beta_{\rm s})
    &= \left(\frac{\nu}{\nu_0}\right)^{\beta_{\rm s}}, \nonumber
\end{align}
corresponding respectively to a MBB SED for thermal dust and a power-law SED for synchrotron emission, both normalized to unity at the reference frequency \(\nu_0\). Evaluating \(A_{\rm cmb}(\nu)\), \(A_{\rm d}(\nu;\beta_{\rm d},T_{\rm d})\), and \(A_{\rm s}(\nu;\beta_{\rm s})\) at each observing frequency and stacking them as columns yields the full mixing matrix \(\mathbf{A}(\boldsymbol{\beta})\).

Assuming Gaussian noise, the negative log-likelihood under this model is:
\begin{equation}
    -2 \ln \mathcal{L}(\mathbf{s}, \boldsymbol{\beta}) = \sum_p (\mathbf{d}_p - \mathbf{A}_p \mathbf{s}_p)^\top \mathbf{N}_p^{-1} (\mathbf{d}_p - \mathbf{A}_p \mathbf{s}_p) + \text{const}.
    \label{eq:nll_joint}
\end{equation}
To estimate the sky components \( \mathbf{s} \), we solve:
\begin{equation}
    \left. \frac{\partial \mathcal{L}}{\partial \mathbf{s}} \right|_{\mathbf{s} = \hat{\mathbf{s}}} = 0,
\end{equation}
which yields the generalized least squares solution:
\begin{equation}
    \hat{\mathbf{s}}_p = \left( \mathbf{A}_p^\top \mathbf{N}_p^{-1} \mathbf{A}_p \right)^{-1} \mathbf{A}_p^\top \mathbf{N}_p^{-1} \mathbf{d}_p \equiv\mathbf{W}\mathbf{d}_p,
    \label{eq:mle_solution}
\end{equation}
Substituting this into the likelihood of Eq.~\eqref{eq:nll_joint} eliminates the dependence on \( \mathbf{s} \), giving the so-called spectral likelihood of \citet{Stompor_2009}:
\begin{equation}
    \ln \mathcal{L}_{\mathrm{spec}}(\boldsymbol{\beta})
    = \mathrm{const}
    + \tfrac{1}{2} \sum_p
    (\mathbf{A}_p^\top \mathbf{N}_p^{-1} \mathbf{d}_p)^\top
    (\mathbf{A}_p^\top \mathbf{N}_p^{-1} \mathbf{A}_p)^{-1}
    (\mathbf{A}_p^\top \mathbf{N}_p^{-1} \mathbf{d}_p), 
    \label{eq:spectral_likelihood}
\end{equation}
which depends only on the spectral parameters \( \boldsymbol{\beta} \) and is used to optimize their values before recovering the sky amplitudes with Eq.~\eqref{eq:mle_solution}.

\subsection{Spatial Modeling via Patching}
\label{subsec:sph_kmeans}

\begin{figure}
    \centering
    \includegraphics[width=\columnwidth]{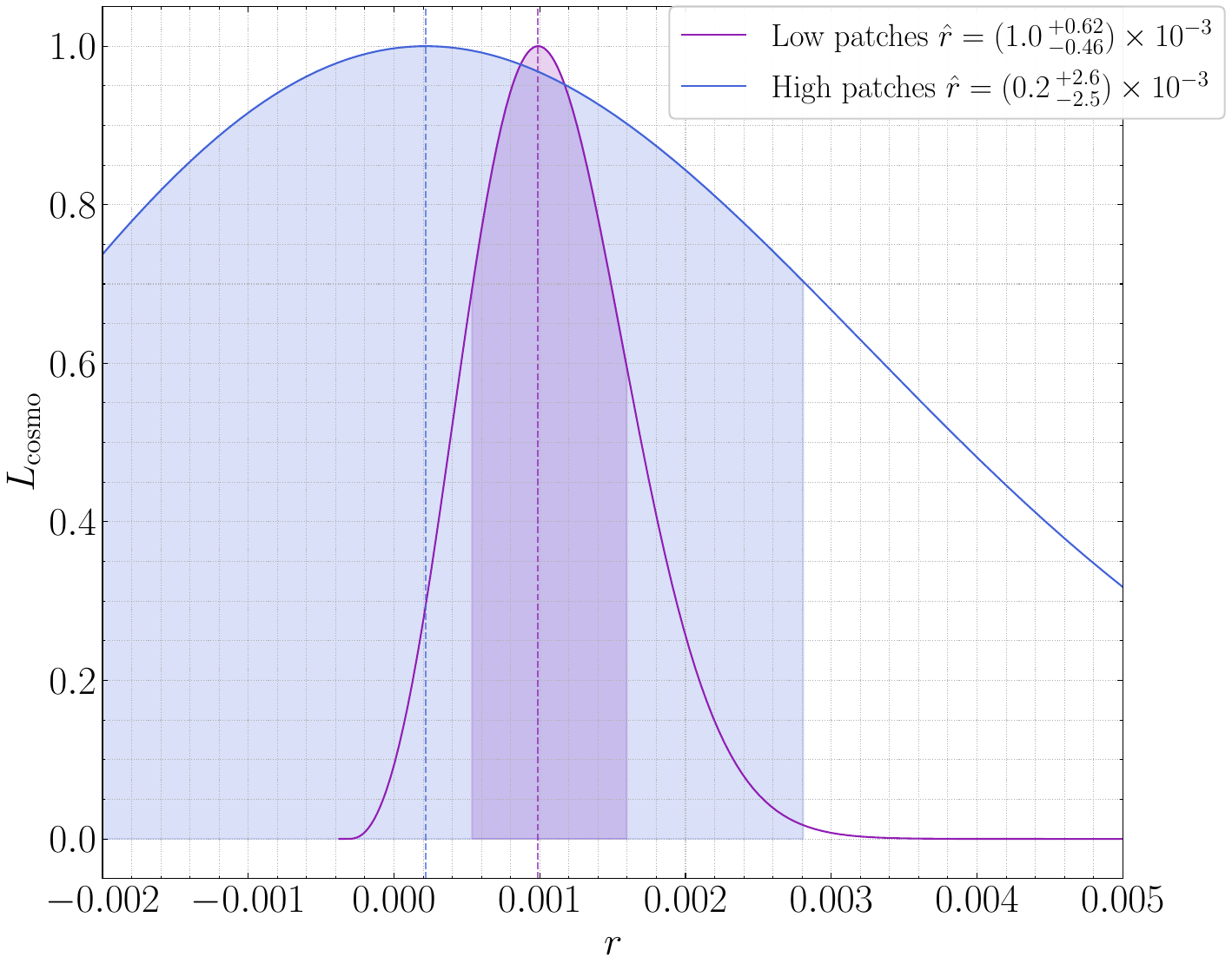}
    \caption{
    Effect on $r$ constraints when considering a low (e.g., $K_{\beta_d} = 4{,}000$, $K_{T_d} = 10$, $K_{\beta_s} = 50$) versus high (e.g., $K_{\beta_d} = 10{,}000$, $K_{T_d} = 3{,}500$, $K_{\beta_s} = 10{,}000$) number of patches to model an input 
     PySM \texttt{c1d1s1} sky with $f_{
    \rm sky} = 60\%$. A low (high) number of patches leads to a biased (unbiased) measurement of the input $r$, associated with low (high) statistical uncertainty; the latter is driven by the number of degrees of freedom. The low-patch case is drawn in purple and the high-patch case in blue; the vertical dashed lines and the shaded bands give the maximum-likelihood $\hat{r}$ and its 68\% interval, and the likelihood is normalized to its peak.}
    \label{fig:illustration_low_high_patches}
\end{figure}

To account for spatial variability in the foreground spectral parameters, we generalize the model of Eq.~\eqref{eq:data_model} by allowing \( \boldsymbol{\beta} \) to vary across the sky. Assigning one parameter set per pixel would excessively increase the statistical uncertainty of the recovered components—particularly the CMB and derived cosmological parameters~\citep{Stompor_2016, Errard_2019}—and would leave larger residuals in the cleaned CMB. \citet{Rizzieri_2025} analyses this trade-off within a parametric framework. Figure~\ref{fig:illustration_low_high_patches} illustrates similar effects.
To mitigate these effects, we introduce a patch-based model in which the sky is divided into patches that share common spectral parameters.

We use our implementation of a spherical K-means algorithm~\citep{Lloyd1982,DhillonModha2001} to partition the sky into disjoint patches \( \{ \mathcal{P}_k \} \), where
\[
\boldsymbol{\beta}_p = \boldsymbol{\beta}_k \quad \text{for all} \quad p \in \mathcal{P}_k .
\]
This assigns a single set of spectral parameters to each patch. The algorithm partitions the sky into roughly equal-area patches, so the chosen number of patches directly defines the characteristic angular size of the patches in the configuration.

\begin{figure}
    \centering
    \includegraphics[width=\linewidth]{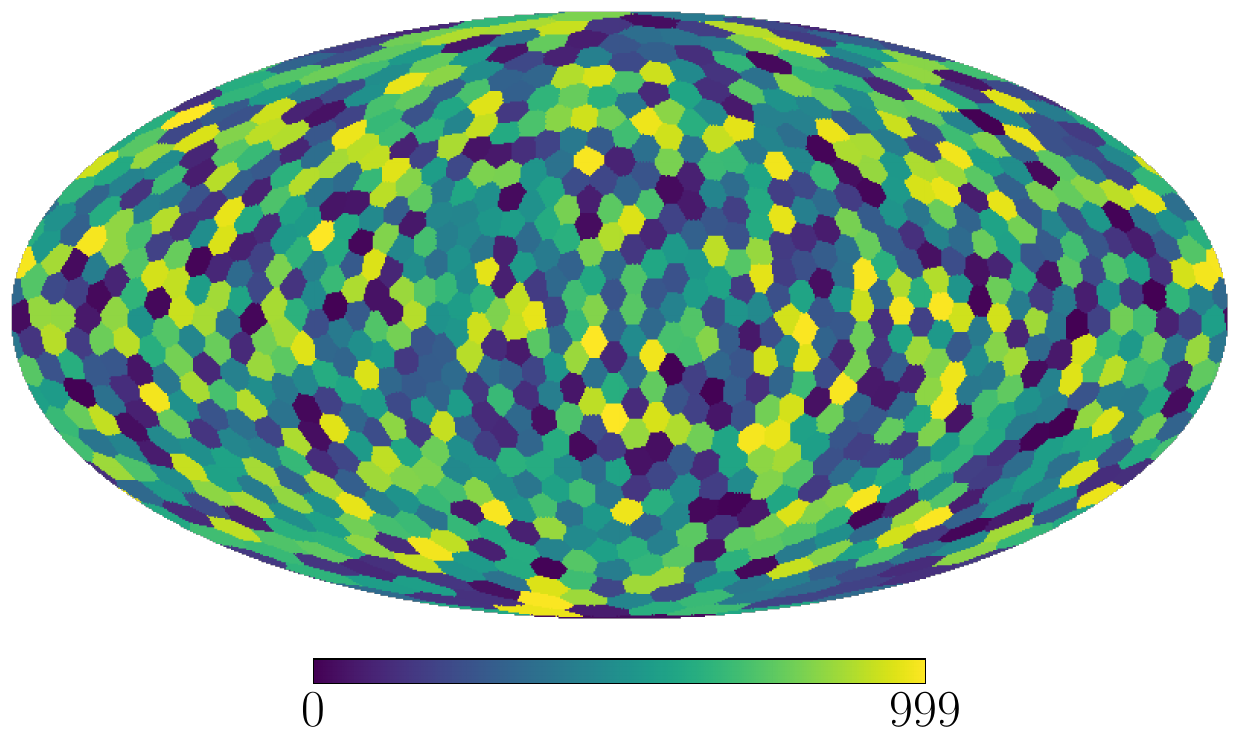}
    \caption{Example of spherical K-means patching applied to a HEALPix-formatted mask \citep{Gorski2005}, showing patches sharing a common spectral parameter. Each color represents a distinct patch.}
    \label{fig:kmeans_clusters}
\end{figure}

Our implementation, available via \texttt{jax-healpy}~\citep{JAXHEALPY}, is adapted for spherical coordinates and inspired by the \texttt{kmeans\_radec} package~\citep{KMEANSRADEC}. The patching algorithm operates on right ascension and declination, and distances are defined as angular separations on the celestial sphere. For numerical stability in the centroid updates, each sky position \((\mathrm{RA}\ \alpha,\mathrm{Dec}\ \delta)\) is first mapped to a unit vector
\[
\hat{\mathbf{n}} = (x,y,z) =
(\cos\delta\cos\alpha,\; \cos\delta\sin\alpha,\; \sin\delta),
\]
and, at each K-means iteration, the centroid of a patch is obtained by averaging these 3D vectors and converting the resulting mean direction back to \((\mathrm{RA},\mathrm{Dec})\). This avoids pathologies due to coordinate singularities and RA wrap-around and leads to robust convergence, as illustrated in Fig.~\ref{fig:kmeans_clusters}.

In practice, to allow different Galactic latitude regions to have their own patch density and characteristic patch size, better adapted to the local foreground complexity, we divide the sky using predefined templates such as the Planck Galactic masks (e.g., \texttt{GAL020}, \texttt{GAL040})\footnote{Available via the Planck Legacy Archive: \url{https://pla.esac.esa.int/}}. These masks divide the sky into regions with different levels of Galactic foreground contamination, typically distinguishing low-, medium-, and high-foreground areas based on thresholding the emission detected by Planck in intensity at 353~GHz.
Each patch configuration—defined by the number of patches used for parameters like \( \beta_{\rm d} \), \( T_{\rm d} \), and \( \beta_{\rm s} \)—specifies a distinct mixing matrix \( \mathbf{A}(\boldsymbol{\beta}) \) within the data model of Eq.~\eqref{eq:data_model}. We emphasize that the way we build one of these K-means patch configurations is not particularly advantageous per se, nor reflecting physical properties of the foregrounds. The power of this approach lays instead in the grid search exploration and the individuation of the resulting optimal configuration, as described in the next section.
\subsection{Grid Search over Patch Configurations}
\label{subsec:grid_search}

Our implementation allows the sky to be patched independently for each spectral parameter (e.g., $\beta_{\rm d},\,T_{\rm d},\,\beta_{\rm s}$), hence allowing for the resulting tilings not to be necessarily spatially aligned across parameters. The numbers of patches assigned to each parameter, $K_{\beta_{\rm d}}, K_{T_{\rm d}}, K_{\beta_{\rm s}}$, are treated as discrete hyperparameters. They are held fixed while optimizing the spectral parameters and recovering the component amplitudes, and are selected externally by a grid search that chooses the configuration minimizing the total error on the recovered tensor-to-scalar ratio $r$ (see Section~\ref{subsec:variance_minimization}). For example, one configuration might use $K_{\beta_{\rm d}}=100$, $K_{T_{\rm d}}=20$, and $K_{\beta_{\rm s}}=10$.

Letting $\mathcal{G}$ denote the space of possible patch counts,
\[
\mathcal{G}=\{K_{\beta_{\rm d}}\}\times\{K_{T_{\rm d}}\}\times\{K_{\beta_{\rm s}}\},
\]
we perform a structured grid search across $\mathcal{G}$. 
For each hyperparameter set $\mathbf{K} \equiv (K_{\beta_d}, K_{T_d}, K_{\beta_s}) \in \mathcal{G}$, we generate the corresponding patch configuration $\mathcal{C}_\mathbf{K}=\left\{ \mathcal{P}_k \right\}$ using the method described in Section~\ref{subsec:sph_kmeans}. We then evaluate this configuration by fitting the spectral parameters $\boldsymbol{\beta}_k$ and reconstructing the CMB component. Specifically, we maximize the spectral likelihood of Eq.~\eqref{eq:spectral_likelihood}, which depends on both the spectral parameter values in each patch and the spatial patch structure:
\begin{equation}
\label{eq:grid_search}
\forall\, \mathcal{C}_\mathbf{K} \in \{\mathcal{C}\} : \quad \boldsymbol{\beta}_k^* = \arg \max_{\boldsymbol{\beta}_k} \, \mathcal{L}_{\mathrm{spec}}(\boldsymbol{\beta}_k, \mathcal{C}_\mathbf{K}),
\end{equation}
where we have denoted the set of all patch configurations evaluated in the grid search as $\{\mathcal{C}\}$. Using these parameters, we then reconstruct the sky components by applying the linear mixing-matrix inversion operator, $\mathbf{W}$, to the data:
\begin{align}
\hat{\mathbf{s}} &= \left( \mathbf{A}(\boldsymbol{\beta}_k^*)^\top \mathbf{N}^{-1} \mathbf{A}(\boldsymbol{\beta}_k^*) \right)^{-1} \mathbf{A}(\boldsymbol{\beta}_k^*) ^\top \mathbf{N}^{-1} \mathbf{d} \label{eq:recon_operator}\\
&\equiv \mathbf{W}(\boldsymbol{\beta}_k^*)\,\mathbf{d},
\end{align}
where $\mathbf{W}$ represents the generalized least-squares solution defined in Eq.~\eqref{eq:mle_solution}. 
From the reconstructed components, we extract the CMB signal \( \mathbf{\hat{s}}_{\mathrm{CMB}} \), and based on that we score the result using the 68\% confidence level limit on $r$ criterion defined in Section~\ref{subsec:variance_minimization}. While maximizing $\mathcal{L}_{\mathrm{spec}}$ ensures a good fit to the multi-frequency data, increasing spatial flexibility can lead to overfitting. The selection metrics introduced in Section~\ref{subsec:variance_minimization} guard against this by favoring configurations with lower residual contamination and improved cosmological performance.

Each configuration is evaluated over multiple noise realizations and sky regions to ensure robustness. Given the large size of $\mathcal{G}$ and the need to sweep many realizations and masks, we use a distributed, parallel evaluation framework described in Section~\ref{subsec:parallel_grid_search}.

\subsection{Selection Metrics for Patch Configurations}
\label{subsec:variance_minimization}

To select the optimal patch configuration, we evaluate metrics that quantify residual contamination and cosmological performance, moving from map-level diagnostics to the final cosmological parameter constraints.

\textbf{Spatial Map Variance ($\sigma^2_{\mathrm{CMB}}$):} 
We first compute the spatial variance of the reconstructed CMB map across the pixels. For a given realization, this is defined as:
\begin{equation}
\label{eq:min_var}
\sigma^2_{\mathrm{CMB}} \equiv \mathrm{Var}_{p \in \mathcal{M}} \left[ \mathbf{\hat{s}}_{\mathrm{CMB}, p} \right],
\end{equation}
where the variance is taken over the pixels $p$ in the analysis mask $\mathcal{M}$. This metric provides a convenient, computationally efficient proxy for the total residual fluctuation (comprising both statistical noise and systematic foreground residuals) in the map. While one could rigorously assess residuals by summing the multipoles of the recovered $B$-mode power spectrum ($\mathcal{P}_{\mathrm{BB}}$), we use the spatial map variance to avoid the computational overhead of explicit $E$-to-$B$ purification on masked maps during rapid grid searches. However, as demonstrated in Section~\ref{subsec:model_selection}, minimizing $\sigma^2_{\mathrm{CMB}}$ alone often favors under-parameterized models that suppress statistical noise but suffer from significant systematic bias (under-fitting of foregrounds).

\textbf{Upper 68\% Confidence Limit on $r$:} 
To explicitly optimize the bias-variance trade-off, our primary selection criterion is the upper limit of the 68\% confidence interval on the tensor-to-scalar ratio, formally $\hat{r} + \sigma(r)$ (see Section~\ref{sec:r_estimation} for the proper likelihood definition), where $\hat{r}$ is the maximum likelihood estimate. This metric explicitly sums the systematic bias (driven by foreground residuals) and the statistical uncertainty. 
Minimizing the maximum likelihood estimate on $\hat{r}$ alone would favor configurations with maximal patch counts (approaching the pixel resolution limit) to capture fine-scale foreground variations, dramatically inflating statistical noise. Minimizing $\sigma(r)$ alone would favor a configuration without patches, likely introducing bias.
Therefore, our final model selection relies on minimizing $\hat{r} + \sigma(r)$, which effectively penalizes both excessive bias and excessive variance.

In principle, these criteria are applicable to both forecasting and real data analysis. The mixing matrix inversion operator $\mathbf{W}$ (Eq.~\ref{eq:recon_operator}) is constructed to possess a unity response to the CMB signal 
\begin{equation}
    \left[ \mathbf{W}\mathbf{A}_{\mathrm{CMB}} \right]_{\mathrm{CMB}} = 1, 
    \label{eq:CMB_unit_response}
\end{equation}
The operator mathematically preserves the CMB spectral response. The reconstructed CMB sky preserves the input CMB signal and adds only statistical noise and systematic foreground residuals. Because the linear solver cannot algebraically destroy the true primordial signal, minimizing the recovered $r$ minimizes the total additive contamination without requiring prior knowledge of the true $r$. Section~\ref{ssec:application_nonzero_r} verifies that this minimization accurately recovers $r$ in simulations with $r_{\text{true}} > 0$. 

When extending this analysis to real data, unmodeled instrumental systematic effects can break this mathematical preservation of the CMB signal. Forthcoming papers will study the impact and mitigation of such systematics.

\subsection{Tensor-to-Scalar Ratio Estimation}
\label{sec:r_estimation}

We estimate the tensor-to-scalar ratio $r$ from the $B$-mode power spectrum of the reconstructed CMB maps. We adopt a maximum likelihood approach.
We use as likelihood on the cosmological parameter $r$ given by the Wishart distribution \citep{WISHART_EIGEN, Gerbino_2020,Hamimeche_2008}, corrected by the effective sky fraction $f_{\mathrm{sky}}$\footnote{Note this is an approximation for partial sky scenarios. This is satisfactory for our proof of concept analysis here as relative comparisons between configurations remain valid, but alternative likelihood approaches should be explored to get reliable cosmological constraints on real data.}:
\begin{equation}
\begin{aligned}
-2\ln\mathcal{L}_{\text{cosmo}}(r)
&= f_{\mathrm{sky}} \sum_{\ell} (2\ell+1)\, \Bigg[
    \frac{C_\ell^{BB,\mathrm{obs}}}{C_\ell^{BB,\mathrm{model}}(r)} \\
&\qquad\qquad + \ln C_\ell^{BB,\mathrm{model}}(r)
\Bigg] + \mathrm{const}.
\end{aligned}
\end{equation}
where the sum runs over the multipole range $\ell_{\min} = 2$ to $\ell_{\max} = 150$. 

In this expression, $C_\ell^{BB,\mathrm{obs}}$ denotes the $B$-mode power spectrum of the reconstructed CMB map.
To avoid $E$-to-$B$ leakage from the masked sky, we subtract the input CMB from the reconstructed maps, leaving the post-component separation residuals. The pseudo-$C_\ell$ of these residuals has negligible E-to-B leakage because the E and B signals in the residuals are of the same order. 
We average the recovered spectra and use them in the likelihood (denoted $C_\ell^\mathrm{tot}$), summed with the input CMB power spectrum.
Approaches exist to avoid this shortcut and deal with the reconstructed CMB map directly, as required for real data, e.g., purification~\citep{Alonso2019}.
We defer such treatment to future work.

We define the model power spectrum as the sum of the CMB primordial tensor signal, the CMB lensing signal, and the average over the simulations of the statistical residuals of the reconstruction:
\begin{equation}
C_\ell^{BB,\mathrm{model}}(r) \;=\; r\,C_\ell^{\mathrm{tensor}} \;+\; C_\ell^{\mathrm{lensing}} \;+\; C_\ell^{\mathrm{stat}},
\end{equation}
where $C_\ell^{\mathrm{tensor}}$ is the primordial tensor template normalized to $r=1$, computed using \textsc{CAMB}~\citep{CAMB}, and $C_\ell^{\mathrm{lensing}}$ is the fixed lensing contribution (corresponding to $A_{\text{lens}}=1$).
The term $C_\ell^{\mathrm{stat}}$ represents the statistical residuals in the reconstructed CMB due to the propagation of instrumental noise through the cleaning process. In our simulation framework, we isolate this contribution by performing a map-level subtraction of the noise-free foreground leakage (the systematic residual) from the total reconstruction error prior to computing the power spectra:
\begin{equation}
C_\ell^{\mathrm{stat}}
\;\equiv\;
\Big\langle C_\ell^{BB}\!\Big(
    \big[\hat{\mathbf{s}}_{\mathrm{CMB}}^{(n)}-\mathbf{s}_{\mathrm{CMB}}^{\mathrm{true}}\big]
    - \big[\mathbf{W}\,\mathbf{d}_{\mathrm{fg}}\big]_{\mathrm{CMB}}
\Big)\Big\rangle_{n},
\label{eq:stat_residuals}
\end{equation}
where $\hat{\mathbf{s}}_{\mathrm{CMB}}^{(n)}$ is the reconstructed CMB for the $n$-th noise realization, $\mathbf{s}_{\mathrm{CMB}}^{\mathrm{true}}$ is the input CMB signal, and the average is taken over noise realizations $n$. 
The term subtracted in the second bracket corresponds to the systematic residuals, which are defined as the leakage of foregrounds into the CMB map in the absence of noise:
\begin{equation}
C_\ell^{\mathrm{syst}} \;\equiv\; C_\ell^{BB}\!\left(\left[\mathbf{W}\,\mathbf{d}_{\mathrm{fg}}\right]_{\mathrm{CMB}}\right).
\label{eq:systematic_residuals}
\end{equation}

When applying this methodology to real data, the true CMB and noise-free foreground maps are unavailable. In such a scenario, $C_\ell^{\mathrm{syst}}$ is not directly accessible and could be misinterpreted as a cosmological signal. 
The exact model for the statistical residuals enables us to perfectly unbias our $r$ estimation from this term.
In a realistic scenario this must be built from the information at our disposal.
Several options have been proposed to estimate the statistical contribution to the foreground residuals, which fundamentally requires estimating the propagation of input noise through the non-linear spectral parameter estimation~\citep{Rizzieri_2025}.
Additionally data splits and a full sampling of the spectral likelihood would help in this noise-debiasing process.

All the residuals introduced here, at power spectrum level, are related through
\begin{equation}
    C_\ell^{\mathrm{tot}} \simeq  C_\ell^{\mathrm{syst}} + C_\ell^{\mathrm{stat}}.
\end{equation}

In this work, we focus on quantifying the level of residual foreground contamination. Since our primary validation simulations assume \( r_{\mathrm{true}} = 0 \), the estimated parameter \( r \) effectively quantifies the bias due to systematic foreground residuals. We therefore interpret the recovered $r$ value as a measure of the contamination level.

\section{Implementation}
\label{sec:implementation}

\subsection{Distributed and Parallel Execution}
\label{subsec:parallel_grid_search}

We developed a distributed optimization engine, \texttt{jax-grid-search}~\citep{kabalan2025jaxgridsearch}, to evaluate large grids of subpixel configurations across multiple noise realizations and sky regions. This framework combines two levels of parallelism to maximize GPU throughput.

We use intra-device parallelism via JAX's \texttt{vmap} transformation to vectorise computations on a single GPU. Specifically, we batch operations over noise realizations, allowing us to solve independent component separation instances simultaneously in a single kernel call, subject only to GPU memory limits.

For inter-device distributed execution, we explore the vast grid of configurations $\mathcal{G}$ by partitioning the workload across multiple processing units (CPUs/GPUs) using a robust chunking strategy. For a grid of size $N_{\mathcal{G}} = |\mathcal{G}|$ and $P$ total processes, each rank $k \in [0, P-1]$ is assigned a contiguous slice of the grid indices $[i_{\mathrm{start}}, i_{\mathrm{end}})$:
\[
i_{\mathrm{start}} = \left\lfloor \frac{k \cdot N_{\mathcal{G}}}{P} \right\rfloor, \quad i_{\mathrm{end}} = \left\lfloor \frac{(k+1) \cdot N_{\mathcal{G}}}{P} \right\rfloor.
\]
This partitioning strategy ensures robust load balancing without assuming that the grid size $N_{\mathcal{G}}$ is perfectly divisible by $P$. It also supports fault tolerance; each batch is evaluated independently, and results are checkpointed to disk, allowing for efficient resumption of interrupted jobs on large HPC clusters.

\subsection{The FURAX Framework}

To support this method, we developed \texttt{FURAX}~\citep{FURAX}, a modular \texttt{JAX}-powered framework to express and optimize parametric component separation likelihoods and CMB data analysis in general.
\texttt{FURAX} is built around a central goal: to implement algebraic operations as composable linear operators directly in code. These operators abstract common structures like the mixing matrix, noise weighting, or parameter dispatching, and can be combined into scalable computation graphs.
\texttt{FURAX} provides:
\begin{itemize}
    \item Differentiable algebraic operators for the \textit{mixing matrix} \( \mathbf{A}(\boldsymbol{\beta}) \), the \textit{noise diagonal} operator \( \mathbf{N}^{-1} \), and cluster-wise \textit{SED parameter dispatch} using mask-based routing.
    \item Conjugate gradient solvers via the \texttt{Lineax} library~\citep{lineax} to solve inverse systems efficiently without explicitly forming matrices.
    \item Fully vectorized execution across simulation ensembles and clustering grids, using \texttt{JAX}~\citep{JAX} for efficient large-scale computation.
\end{itemize}

\begin{figure}
    \centering
    \input{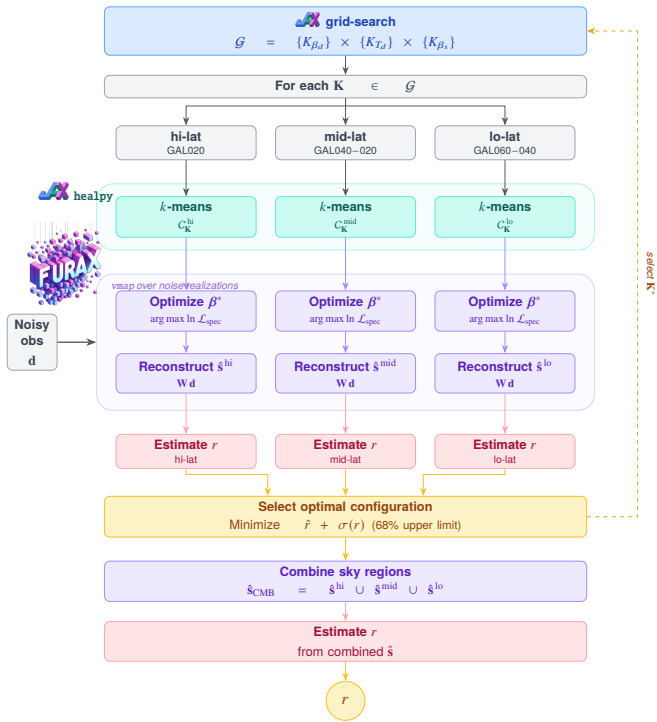}
    \caption{Overview of the end-to-end pipeline for adaptive component separation and tensor-to-scalar ratio estimation.
        The grid search (top, blue) enumerates all configurations $\mathbf{K} \in \mathcal{G}$ (Section~\ref{subsec:grid_search}), distributed across devices via \texttt{jax-grid-search}~\citep{kabalan2025jaxgridsearch}.
        For each configuration, the pipeline operates independently on three disjoint sky regions (high-latitude (\emph{hi-lat}, GAL020), mid-latitude (\emph{mid-lat}, GAL040$\,-\,$020), and low-latitude (\emph{lo-lat}, GAL060$\,-\,$040)), defined by Planck Galactic masks (see Section~\ref{subsec:sky_regions} and Fig.~\ref{fig:mask_layout}).
        Within each region, spherical K-means clustering (\texttt{jax-healpy}~\citep{JAXHEALPY}, teal container) partitions the sky into a patch configuration $\mathcal{C}_{\mathbf{K}}$, spectral parameters $\boldsymbol{\beta}^*$ are optimized by maximizing $\mathcal{L}_{\mathrm{spec}}$ (Eq.~\ref{eq:spectral_likelihood}), and CMB maps are reconstructed via $\hat{\mathbf{s}} = \mathbf{W}\,\mathbf{d}$ (Eq.~\ref{eq:recon_operator}), all within the \texttt{FURAX} framework (purple container), with the optimization and reconstruction vectorized over noise realizations via \texttt{vmap}.
        The tensor-to-scalar ratio $r$ is then estimated independently for each region at $f_{\mathrm{sky}} = 20\%$ (coral), and the configuration minimizing $\hat{r} + \sigma(r)$ (68\% upper limit, Section~\ref{subsec:variance_minimization}) is selected (amber).
        Post-selection, the three optimally cleaned regional CMB maps are combined into a joint map covering $f_{\mathrm{sky}} = 60\%$, from which the final estimate of $r$ is derived with reduced uncertainty.
        }

    \label{fig:furax_pipeline}
\end{figure}

Figure~\ref{fig:furax_pipeline} shows the end-to-end workflow of our component separation pipeline, built on the \texttt{FURAX} framework. A patch configuration is selected from the discrete grid $\mathcal{G}$, spectral parameters are optimized by maximizing $\mathcal{L}_{\mathrm{spec}}$, maps are reconstructed via the linear operator $\mathbf{W}$ (Eq.~\ref{eq:recon_operator}), and selection metrics (CMB variance, summed B-mode power, and the $r$-based metric; see Section~\ref{subsec:variance_minimization}) identify the best configuration before estimating $r$ from the recovered B modes.

All core computations in FURAX use memory-efficient linear operators, avoiding explicit dense matrices. This symbolic and modular approach allows \texttt{FURAX} to express and optimize large, realistic models without the memory bottlenecks of explicit matrix construction. This is crucial for next-generation experiments like LiteBIRD~\citep{LITEBIRD_PTEP_2022} where the time-domain data volume prohibits explicit matrix storage. Several companion works extending \texttt{FURAX} to additional instrument modelling capabilities are in preparation, with one already submitted dealing with the modelling of the half-wave plate~\citep{EMA}.

\texttt{FURAX} accommodates more realistic data models, including beam convolution, correlated noise, and instrument-specific systematics, by composing additional linear operators into the existing pipeline. In this paper, we focus on a simplified validation setup with isotropic white noise at low HEALPix~\citep{Gorski2005} resolution ($N_{\text{side}} = 64$), with no explicit beam convolution applied (all frequency channels are assumed to share a common effective resolution). At this resolution, beam effects are subdominant to foreground contamination. Future work will extend the analysis to higher resolution and more complex noise/beam \citep{sathyanathan2025parameterizing, rizzieri2025validating} models.

The framework integrates natively with the \texttt{JAX} ecosystem and supports optimization and inference workflows through libraries such as \texttt{Optax}~\citep{optax} and \texttt{NumPyro}~\citep{numpyro}.

\subsection{High-Performance Spectral Likelihood maximization}
\label{sec:spectral_likelihood_maximization}

The central task in our parametric component separation is estimating the spectral parameters \( \boldsymbol{\beta} \) by maximizing the spectral likelihood \( \mathcal{L}_{\mathrm{spec}}(\boldsymbol{\beta}) \) (Eq.~\eqref{eq:spectral_likelihood}). This likelihood drives parameter inference across the model.

Identifying the optimal bias-variance trade-off requires evaluating thousands of patch configurations across tens of noise realizations (40 in our production runs). This scale demands an optimizer significantly faster than standard CPU implementations. This section presents our optimized differentiation strategy and the custom \texttt{AdaTopK} optimization method, which outperforms the standard \texttt{scipy} TNC optimizer in both speed and solution quality.
\subsubsection{Spectral Likelihood Gradient Implementation}

We evaluate the spectral likelihood using the \texttt{FURAX} framework. The central computational step is the inversion of the curvature matrix \(\mathbf{M} = \mathbf{A}^\top \mathbf{N}^{-1} \mathbf{A}\) to solve for the sky components \(\mathbf{s}\).

Minimizing the likelihood requires computing gradients with respect to the spectral parameters \(\boldsymbol{\beta}\). In standard reverse-mode automatic differentiation, differentiating through an iterative linear solver typically relies on implicit differentiation. This involves solving an intermediate linear system to apply the chain rule and calculate how changes in the parameters affect the final output. Naive application of this method can cause numerical instabilities when the forward operator \(\mathbf{A}(\boldsymbol{\beta})\) embeds strict physical constraints (such as the strict positivity of the dust temperature). During the intermediate steps of automatic differentiation, the algorithm may attempt to evaluate gradients at non-physical parameter values that violate these domain constraints, triggering errors or instabilities when computing the gradient using JAX's automatic differentiation.

To resolve this, we implement a custom backward pass using \texttt{jax.custom\_vjp} that bypasses the implicit differentiation of the solver entirely. Instead, we use the analytical gradient of the concentrated spectral likelihood. We explicitly compute the data residuals \(\mathbf{w}_r = \mathbf{N}^{-1}(\mathbf{d} - \mathbf{A}\mathbf{s})\) using the valid, physical sky components \(\mathbf{s}\) recovered in the forward pass. By exploiting the condition that \(\mathbf{s}\) minimizes the generalized least squares objective, the gradient of the spectral likelihood \(\mathcal{L}_{\mathrm{spec}}\) with respect to the parameters can be expressed solely in terms of these residuals:
\begin{equation}
    \nabla_{\boldsymbol{\beta}} \mathcal{L}_{\mathrm{spec}} = 2 \mathbf{w}_r^\top \nabla_{\boldsymbol{\beta}} \big( \mathbf{A}(\boldsymbol{\beta})\,\mathbf{s} \big),
\end{equation}
where the notation indicates that the sky components are treated as fixed constants for the purpose of this derivative. This formulation ensures that the gradient computation depends only on the mixing matrix \(\mathbf{A}\) and its derivatives evaluated at the current, physically valid parameter point \(\boldsymbol{\beta}\), guaranteeing numerical stability throughout the optimization. While we bypass the solver's implicit differentiation, standard JAX automatic differentiation still evaluates \(\nabla_{\boldsymbol{\beta}} \mathbf{A}(\boldsymbol{\beta})\), handling complex operations such as the routing of cluster parameters to their corresponding individual pixels.

\subsubsection{The AdaTopK Optimizer}

Minimizing the spectral likelihood for thousands of sub-pixel clusters is a high-dimensional, bounded optimization problem. Standard quasi-Newton methods, such as L-BFGS~\citep{Liu1989} or Truncated Newton (TNC)~\citep{TNC}, often struggle in this regime. In particular, they perform poorly in low-SNR regions where gradients are dominated by noise. The TNC algorithm detects unreliable gradient steps and responds by resetting its L-BFGS curvature history, which stabilizes the minimization process but significantly slows convergence in noisy regimes.

We developed \textbf{AdaTopK} (Adaptive AdaBelief with Top-K Active Set), a custom JAX-native optimizer that reimplements the TNC active-set constraint strategy natively in JAX. The optimizer maintains an active set of bound-constrained parameters and releases them based on gradient alignment with the feasible direction. A Top-K fraction hyperparameter $K$ controls how many constraints are released per iteration; after systematic evaluation (see Appendix~\ref{app:topk_impact}), we adopt $K=0$ (single-parameter release, matching TNC behaviour), which provides the most robust convergence.

AdaTopK offers three advantages over the \texttt{scipy} TNC implementation: (1) being JAX-native, it runs entirely on the GPU, enabling efficient vectorization over all grid points via \texttt{jax.vmap}; (2) the underlying AdaBelief optimizer~\citep{Zhuang2020AdaBelief} is better suited to the noisy gradient landscape of low-SNR regions than classical quasi-Newton updates; and (3) a dynamic state rescaling mechanism keeps gradients within a numerically safe range across the extreme dynamic range between the bright Galactic plane and the faint high-latitude sky, without resetting the optimizer's momentum.

\textit{For the full algorithmic details, including the internal parameter mapping, release score definitions, and rescaling equations, see Appendix~\ref{app:adatopk}.}

\subsubsection{Performance Benchmarks}

\begin{figure}
    \centering
    \includegraphics[width=\linewidth]{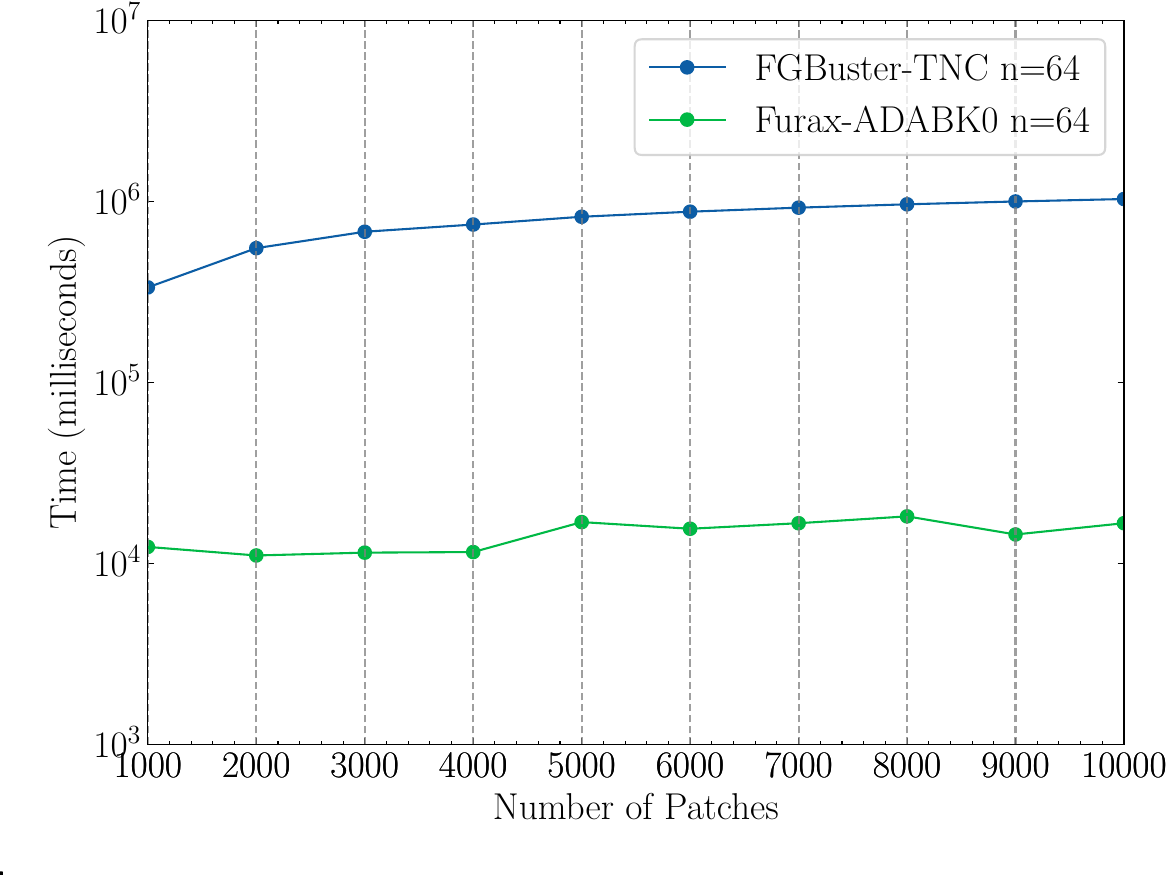}
    \caption{Runtime comparison of the spectral likelihood minimization for AdaTopK (this work) versus \texttt{scipy} TNC (CPU). The x-axis shows the total number of spectral parameter patches, distributed as 50\% $\beta_d$, 30\% $T_d$, and 20\% $\beta_s$. The y-axis shows the wall-clock time in milliseconds to reach convergence. AdaTopK on GPU converges up to ${\sim}100\times$ faster than TNC on CPU at $10{,}000$ patches, with the speed-up reflecting both algorithmic improvements and CPU-to-GPU hardware parallelism.}
    \label{fig:runtime_comparison}
\end{figure}

We benchmark AdaTopK against the standard \texttt{scipy}~\citep{scipy} TNC implementation, which gives satisfactory results with \textsc{FGBuster}~\citep{Rizzieri_2025}. The test was performed on a masked sky ($f_{\rm sky} = 60\%$, $N_{\text{side}}=64$) with $\approx 33{,}000$ total parameters, distributed as 50\% $\beta_d$, 30\% $T_d$, and 20\% $\beta_s$ patches. AdaTopK was benchmarked on a single NVIDIA H100 GPU, while TNC ran on a single \textit{Jean Zay} CPU node with 40 cores. Although the TNC optimizer itself executes serially, including the active-set pivot updates described in Appendix~\ref{app:adatopk}, the underlying linear algebra operations (matrix products, Hessian evaluations) benefit from multi-threaded BLAS on the available cores; the comparison reflects a realistic deployment scenario for each platform.
Figure~\ref{fig:runtime_comparison} illustrates the convergence speed as a function of the total number of parameter patches. AdaTopK on GPU converges up to ${\sim}100\times$ faster than TNC on CPU at $10{,}000$ patches, demonstrating the combined advantage of algorithmic improvements and GPU-accelerated optimization.

Beyond raw speed, AdaTopK shows improved robustness at avoiding false local minima compared with the TNC-based \textsc{FGBuster} implementation. In our tests, TNC frequently converges to suboptimal solutions, particularly in low signal-to-noise regions where gradients are dominated by noise. This manifests as parameters becoming locked at their bounds and, more critically, entire patches settling into false minima from which the quasi-Newton update cannot escape. AdaTopK mitigates this through three complementary mechanisms: (1) the AdaBelief optimizer~\citep{Zhuang2020AdaBelief} tracks the variance of gradient deviations, naturally adapting step sizes in noisy regions where TNC's Hessian approximation becomes unreliable; (2) the dynamic state rescaling (Appendix~\ref{app:adatopk}) prevents numerical stalling across the extreme dynamic range of the sky signal, continuously rescaling both the objective and the optimizer's internal moments; and (3) a backtracking line search ensures each step yields an actual decrease in the objective, preventing momentum-driven overshooting into worse basins.

\subsection{Sky Region Partitioning}
\label{subsec:sky_regions}

\begin{figure}
    \centering
    \includegraphics[width=0.8\linewidth]{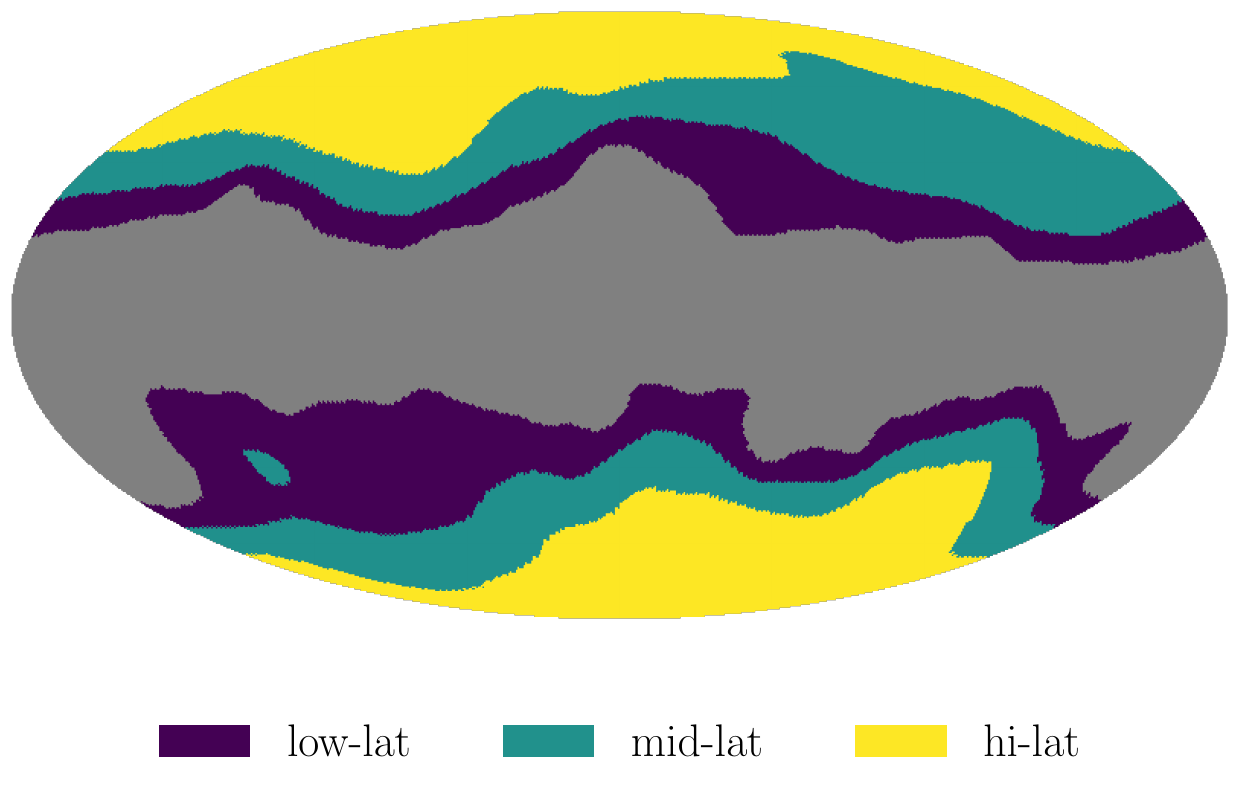}
    \caption{Sky partitioning into three disjoint regions based on the Planck Galactic plane masks~\citep{Planck2018IV}. The high-latitude region retains the cleanest 20\% of the sky ($f_{\rm sky} = 0.2$, GAL020, hereafter \emph{hi-lat}), the mid-latitude region covers the next 20\% ($0.2 < f_{\rm sky} \leq 0.4$, hereafter \emph{mid-lat}), and the low-latitude region the following 20\% ($0.4 < f_{\rm sky} \leq 0.6$, hereafter \emph{low-lat}). The component separation pipeline is applied independently to each region. Hence, the total $f_{\rm sky}$ used for the analysis is 60\%.}
    \label{fig:mask_layout}
\end{figure}

To give the algorithm more flexibility and better adaptability to the input sky complexity, we assessed performance across different foreground regimes by partitioning the sky into several disjoint regions, as illustrated in Figure~\ref{fig:mask_layout}. This partitioning allows spatially adaptive regularization: different regions exhibit vastly different signal-to-noise ratios and foreground complexities. Running the component separation independently on these sub-regions lets us optimize the patch configuration (i.e., the number of clusters) separately for each mask. Low Galactic latitudes, characterized by high foreground amplitudes and complex spatial variations, require a finer patch structure (higher patch density) to minimise modelling bias. High-latitude regions, where the foreground signal is faint and noise-dominated, benefit from larger patches to average down statistical noise.

We hence define three regions based on the Planck Galactic masks~\citep{Planck2018IV}:
\begin{itemize}
    \item \textbf{High-latitude region (hi-lat)}: Defined by the \texttt{GAL020} mask.
    \item \textbf{Mid-latitude region (mid-lat)}: Between \texttt{GAL040} and \texttt{GAL020}.
    \item \textbf{Low-latitude region (low-lat)}: Between \texttt{GAL060} and \texttt{GAL040}.
\end{itemize}

\section{Results}
\label{sec:results}

We validate the pipeline on simplified synthetic data (Section~\ref{subsec:validation}), apply it to realistic \texttt{PySM3} \texttt{d1s1} simulations (Section~\ref{subsec:model_selection}), compare it with the multi-resolution baseline (Section~\ref{subsec:kmeans_vs_multires}), test it on a sky with non-zero $r$ (Section~\ref{ssec:application_nonzero_r}), and explore coarser partitions by binning optimized parameters (Section~\ref{ssec:post_clustering}). All noise realisations use white noise with $\mu$K-arcmin as given in~\cite{LITEBIRD_PTEP_2022}.

\subsection{Validation on Simplified Synthetic Data}
\label{subsec:validation}

We first validate the pipeline on two controlled setups where the ground-truth spectral configuration is known exactly. In both cases we use the full nominal noise level and scan $K_{\beta_d}$ from 1 to 300 (step 10), selecting the configuration that minimizes the 68\% upper limit on $r$ (i.e.\ $r + \sigma(r)$).

The first case uses PySM's \texttt{c1d0s0} sky, which has spatially uniform dust and synchrotron spectral parameters. Because the SEDs do not vary across the sky, a single patch per parameter suffices; we therefore fix $K_{T_d}=1$ and $K_{\beta_s}=1$ and scan only $K_{\beta_d}$.
The second case uses a synthetic sky constructed on a K-means grid with $K_{\beta_d}=100$, $K_{T_d}=15$, and $K_{\beta_s}=5$, so that the true optimal configuration is known by construction. We again scan $K_{\beta_d}$ while keeping $K_{T_d}$ and $K_{\beta_s}$ at their ground-truth values.

\begin{figure}
    \centering
    \includegraphics[width=\columnwidth]{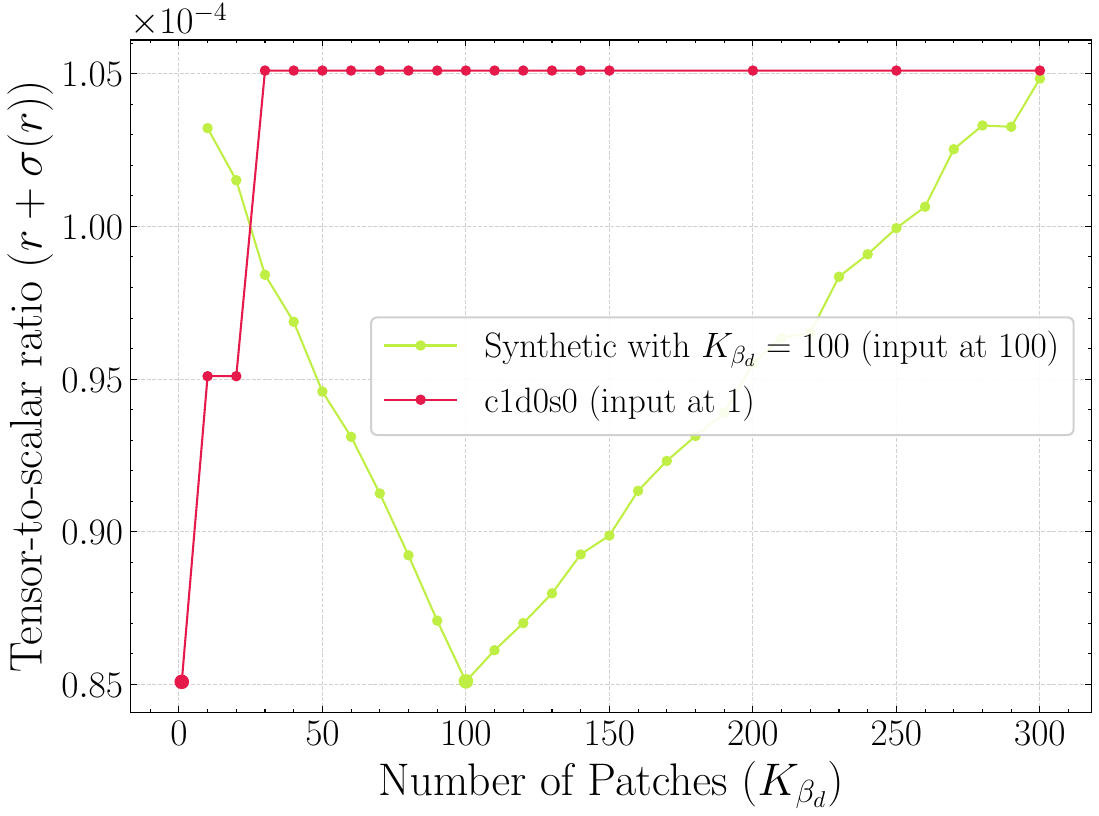}
\caption{
Validation on simplified synthetic data. The 68\% upper limit on $r$ ($r + \sigma(r)$) is plotted as a function of the number of dust spectral-index patches $K_{\beta_d}$. Green: synthetic sky with $K_{\beta_d}=100$, $K_{T_d}=15$, $K_{\beta_s}=5$; red: \texttt{c1d0s0} sky (uniform SEDs). Dashed vertical lines mark the ground-truth $K_{\beta_d}$ for each case. The $y$-axis is truncated at $1.05 \times 10^{-4}$; configurations with $r + \sigma(r)$ above this threshold are clipped, producing the apparent ceiling visible for $K_{\beta_d} \gtrsim 30$.}
    \label{fig:validation}
\end{figure}

Fig.~\ref{fig:validation} shows the results. For the \texttt{c1d0s0} sky, the metric increases sharply as soon as patches are added and then plateaus: a uniform sky is correctly identified as not needing additional patches. For the synthetic $K_{\beta_d}$=100 sky, the curve is V-shaped with a clear minimum at $K_{\beta_d}=100$, confirming that the pipeline recovers the true input configuration.

\subsection{Spatially Varying SEDs}
\label{subsec:model_selection}

We now apply our pipeline to the realistic \texttt{c1d1s1} PySM3 sky model with spatially varying dust and synchrotron spectral parameters, combined with realistic LiteBIRD noise levels.
We perform a grid search (Section~\ref{subsec:grid_search}) over clustering configurations \( (K_{\beta_d}, K_{T_d}, K_{\beta_s}) \), scanning each parameter over the 24 values listed in Table~\ref{tab:grid_config}.
The pipeline is applied independently to three Galactic regions (Section~\ref{subsec:sky_regions}): hi-lat (GAL020), mid-lat (GAL040$-$GAL020), and low-lat (GAL060$-$GAL040).
A single CMB realization is used throughout; cosmic variance across CMB realizations is not explored in this work. For each configuration, 40 noise realizations are evaluated, and the optimal patch counts are selected by minimizing the 68\% upper limit on \( r \) (Section~\ref{subsec:variance_minimization}).
Computations are carried out on the \textit{Jean Zay} supercomputer\footnote{\url{https://www.idris.fr/jean-zay/jean-zay-presentation.html}} using NVIDIA H100 GPUs, with 64 parallel jobs completing in approximately 4 hours of wall-clock time ($\sim$256 GPU-hours total).

\begin{table}
    \centering
    \begin{tabular}{c|l}
        Parameter & Values \\
        \hline
        $K_{\beta_d}$, $K_{T_d}$, $K_{\beta_s}$ & 50, 100, 200, 300, 500, 1000, 1500, 2000, 2500, 3000, \\
                                                  & 3500, 4000, 4500, 5000, 5500, 6000, 6500, 7000, 7500, \\
                                                  & 8000, 8500, 9000, 9500, 10000 \quad (24 values each) 
    \end{tabular}
    \caption{Grid search values for each spectral parameter. The same 24 patch counts are used for all three parameters. When the requested patch count exceeds the number of unmasked pixels in a region (${\sim}9{,}800$ at $N_{\text{side}}=64$ for $f_{\text{sky}}=20\%$), each pixel is assigned its own patch (see Section~\ref{subsec:sky_regions}). The same grid values are used for all three Galactic regions.}
    \label{tab:grid_config}
\end{table}

Our analysis reveals a relationship between the variance of the recovered CMB map and the bias on the tensor-to-scalar ratio \( r \). For the noise levels considered, statistical residuals dominate systematic ones for all but the smallest patch counts, so the spatial variance of the recovered map (summing the $Q$ and $U$ Stokes parameters) proxies the \emph{statistical} error on \( r \). Minimizing this variance drives the model toward few patches, which under-fit the foregrounds and \emph{increase} the systematic bias (recall Fig.~\ref{fig:illustration_low_high_patches}); minimizing \( r \) alone favors the maximum number of patches and inflates statistical noise. Variance alone is insufficient; the optimal configuration balances statistical error against systematic bias.

\begin{figure}
    \centering
    \includegraphics[width=1.05\columnwidth]{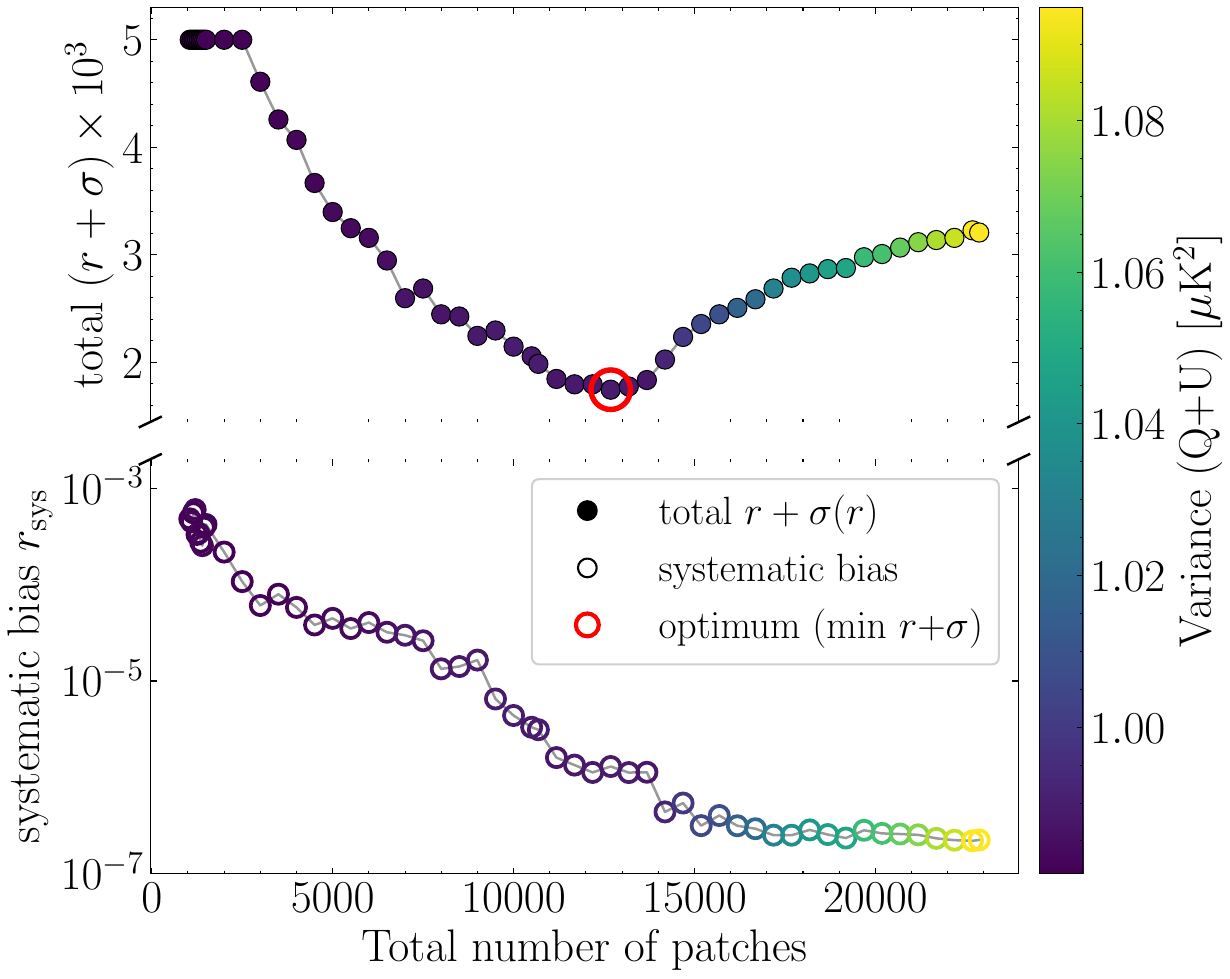}
    \caption{Bias--variance trade-off along a chain of monotonically increasing model complexity (low-latitude region GAL060). The patch counts are raised in three successive phases: first the dust spectral index $K_{\beta_d}$ ($50\!\rightarrow\!\mathrm{max}$, at $K_{T_d}=K_{\beta_s}=500$), then the dust temperature $K_{T_d}$ (up to $3500$, at $K_{\beta_d}=\mathrm{max}$, $K_{\beta_s}=500$), and finally the synchrotron spectral index $K_{\beta_s}$ (up to max, at $K_{\beta_d}=\mathrm{max}$, $K_{T_d}=3500$), so the total patch count ($x$-axis) grows throughout. As the foregrounds are progressively resolved the systematic bias decreases (lower panel, log scale) while the recovered-map variance rises (color); their sum, the $68\%$ upper limit $r+\sigma(r)$ (upper panel), traces a clean V whose minimum (red ring) marks the optimal trade-off between bias and statistical error. Fig.~\ref{fig:r_vs_clusters_seq} shows the per-parameter projections.}
    \label{fig:variance_vs_r}
\end{figure}

The 68\% upper limit on \( r \), defined as \( r + \sigma(r) \), combines systematic bias and statistical uncertainty into a single metric. Figure~\ref{fig:variance_vs_r} shows a clear V-shaped behaviour, revealing an optimal number of patches that balances variance and bias and minimizes the total error budget. We find and characterise this optimal configuration.

\begin{figure*}
    \centering
    \begin{minipage}[c]{0.42\textwidth}
        \begin{subfigure}{\textwidth}
            \centering
            \includegraphics[width=\linewidth]{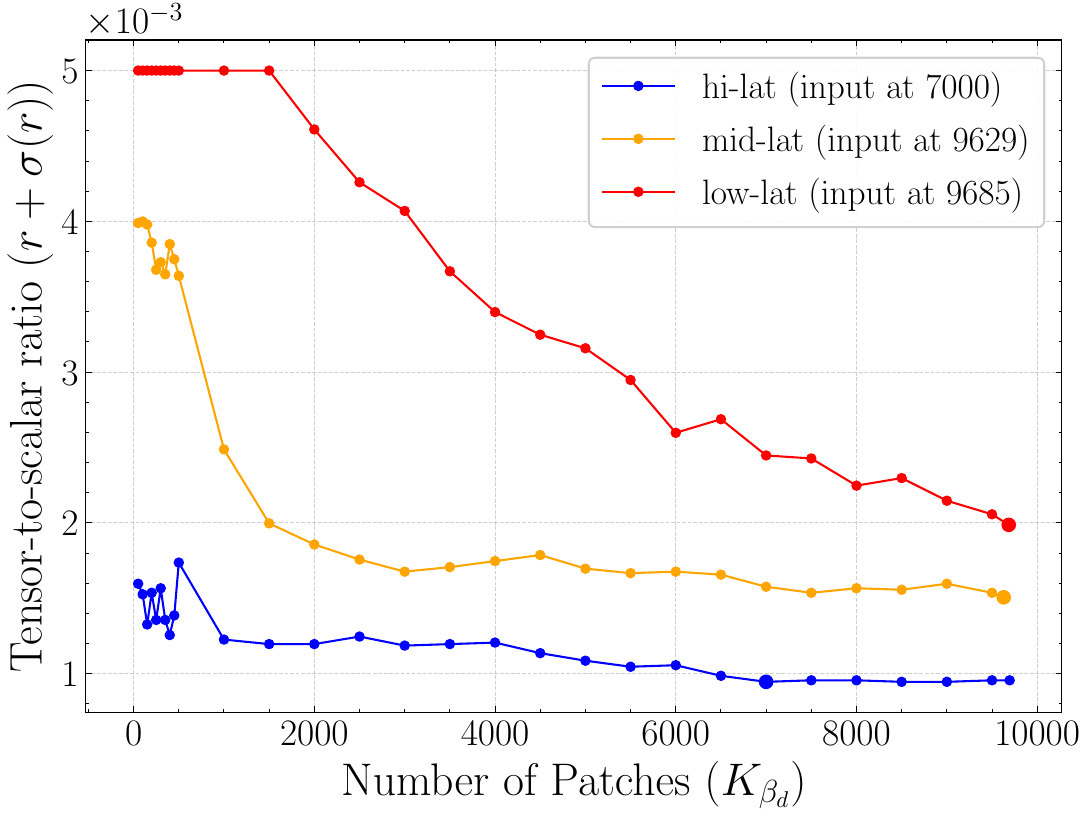}
            \caption{Varying \( K_{\beta_d} \) ($K_{T_d}=500$, $K_{\beta_s}=500$)}
            \label{fig:r_vs_bd}
        \end{subfigure}
        \vspace{1em} 

        \begin{subfigure}{\textwidth}
            \centering
            \includegraphics[width=\linewidth]{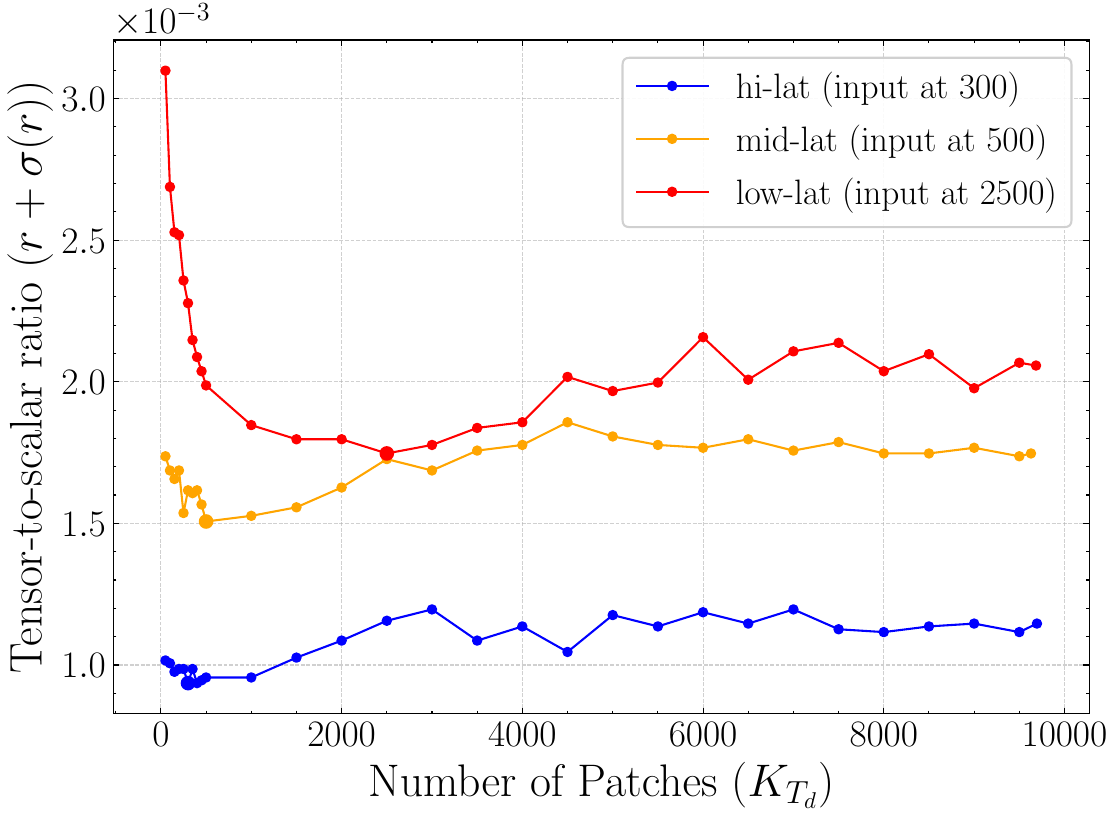}
            \caption{Varying \( K_{T_d} \) ($K_{\beta_d}=\,$max, $K_{\beta_s}=500$)}
            \label{fig:r_vs_td}
        \end{subfigure}
        \vspace{1em}

        \begin{subfigure}{\textwidth}
            \centering
            \includegraphics[width=\linewidth]{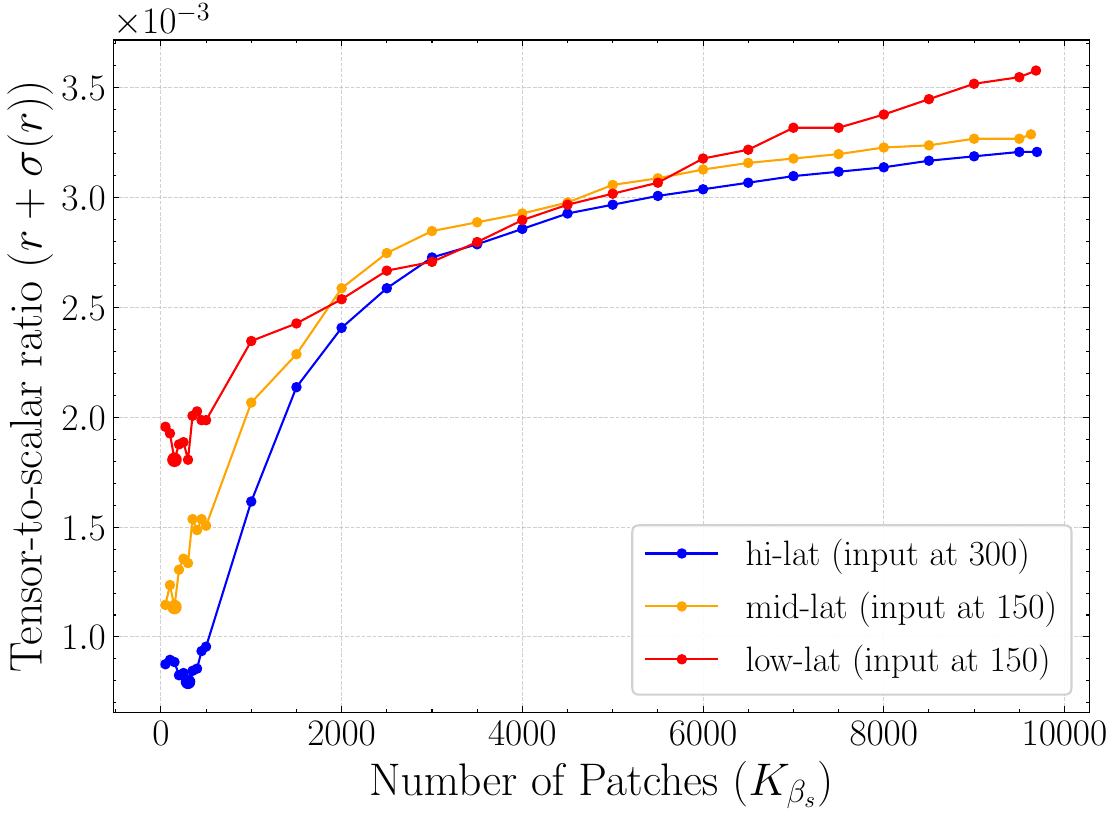}
            \caption{Varying \( K_{\beta_s} \) ($K_{\beta_d}=\,$max, $K_{T_d}=500$)}
            \label{fig:r_vs_bs}
        \end{subfigure}
    \end{minipage}\hfill
    \begin{minipage}[c]{0.54\textwidth}
        \begin{subfigure}{\textwidth}
            \centering
            \includegraphics[width=\linewidth]{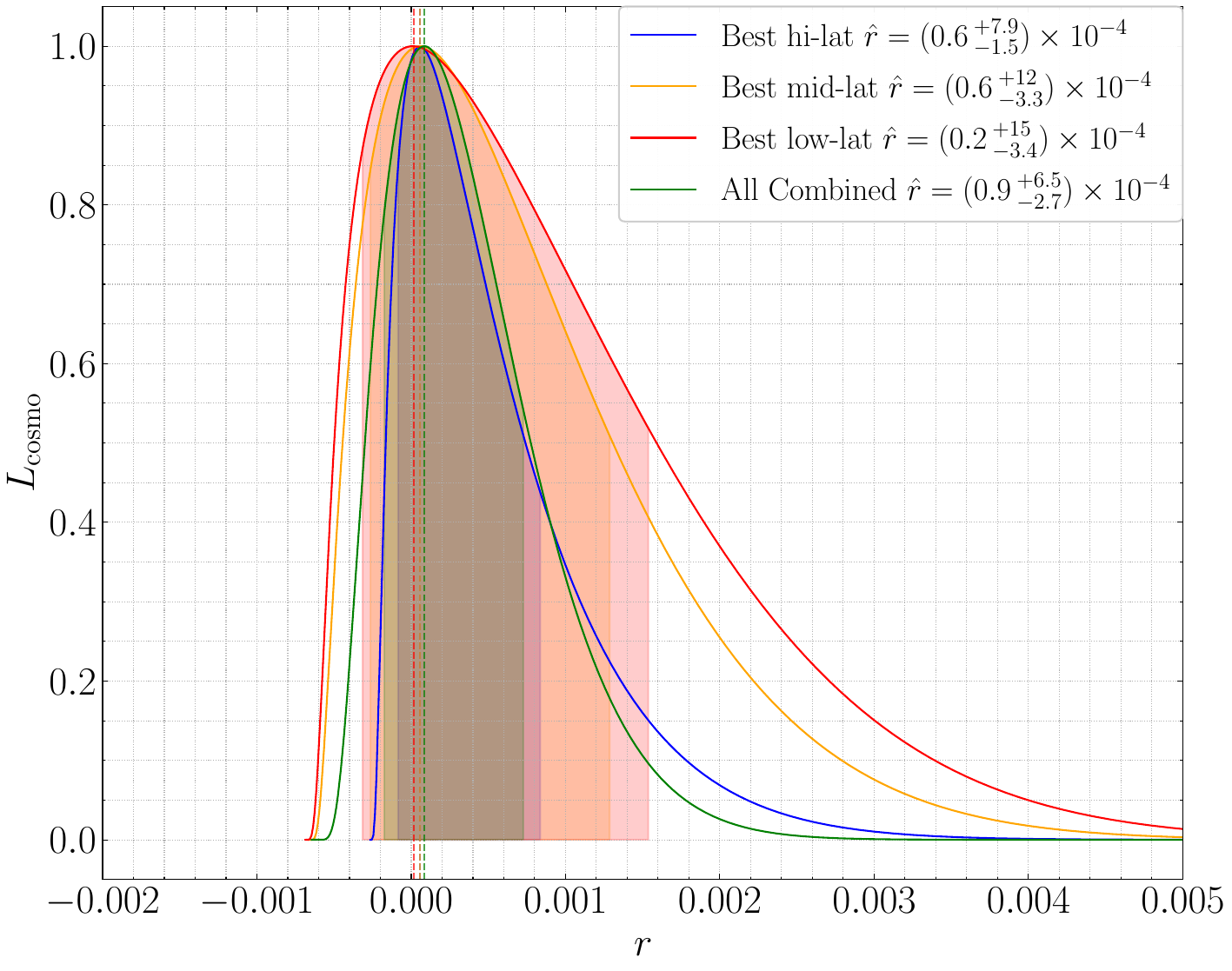}
            \caption{Posterior likelihood on $r$ for the optimal configuration of each region and their combination. Dashed lines and shaded bands indicate the maximum-likelihood $\hat{r}$ and 68\% confidence interval.}
            \label{fig:r_likelihood_optimal}
        \end{subfigure}
        \vspace{3em} 

        \begin{subfigure}{\textwidth}
            \centering
            \includegraphics[width=\linewidth]{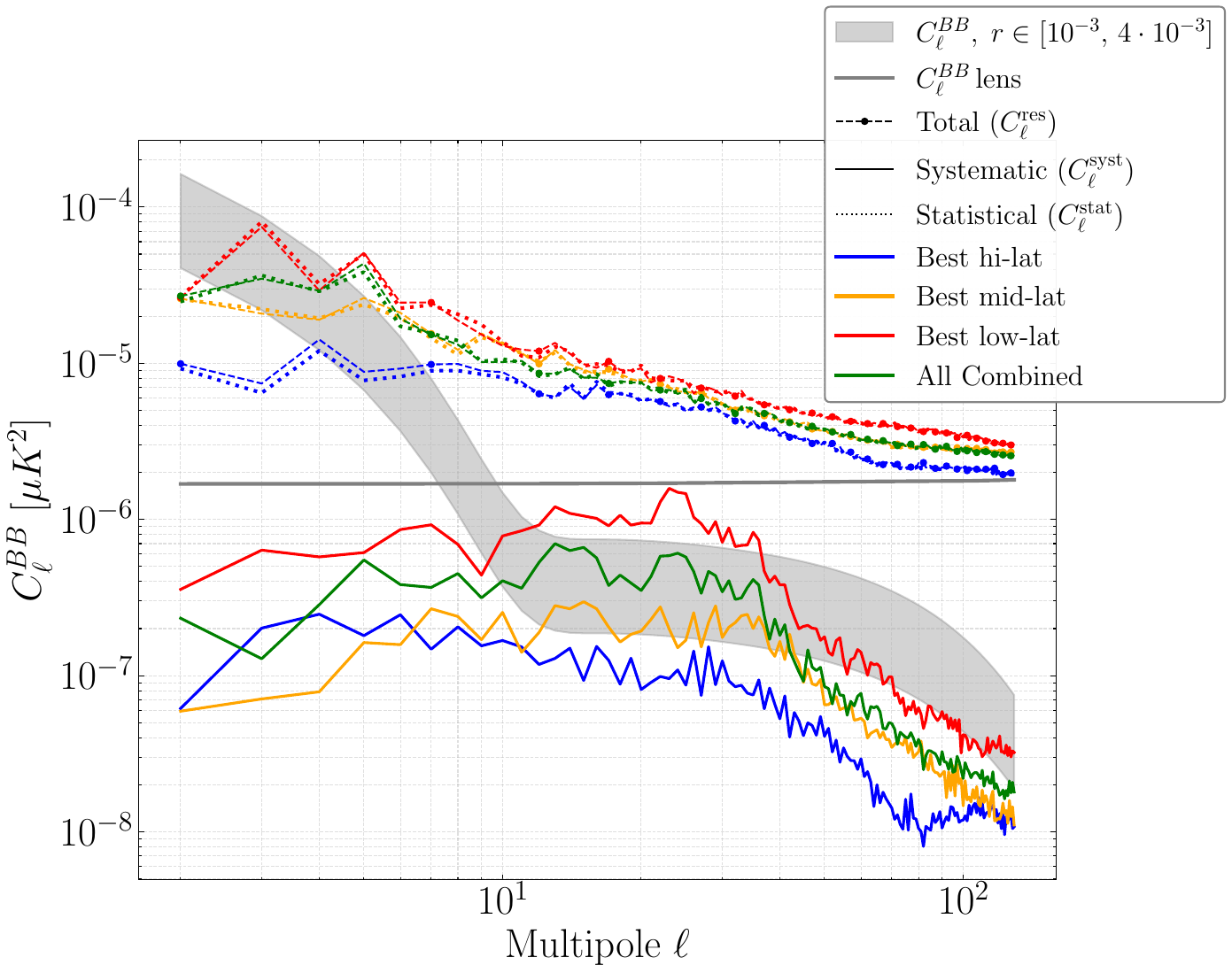}
            \caption{Residual $B$-mode power spectra for the optimal configuration: total (dashed), systematic (solid), and statistical (dotted) residuals. The grey band shows the primordial $C_\ell^{BB}$ for $r \in [10^{-3}, 4 \times 10^{-3}]$; the solid grey line is the lensing contribution.}
            \label{fig:bb_spectra_optimal}
        \end{subfigure}
    \end{minipage}

    \caption{Grid search results on the \texttt{c1d1s1} sky. Left column: 68\% upper limit on $r$ ($r + \sigma(r)$) as a function of the number of patches for each spectral parameter, shown for the three Galactic regions (hi-lat, mid-lat, low-lat). The other two parameters are fixed at representative values (see text). Right column: resulting $B$-mode power spectra for the systematic, statistical and total residuals averaged over the number of realizations (bottom) and posterior likelihood on $r$ (top) for the optimal configuration in each region.}
    \label{fig:r_vs_clusters_seq}
\end{figure*}

Figure~\ref{fig:r_vs_clusters_seq} shows 1D projections of the full grid: for each spectral parameter, we plot the 68\% upper limit on $r$ as a function of that parameter's patch count while fixing the other two at representative values ($K_{T_d}=500$ and $K_{\beta_s}=500$). The optimal values in Table~\ref{tab:optimal_clusters_results} come from the full $24^3$ grid search over all three parameters jointly, evaluated independently for each Galactic region. Three distinct behaviours emerge.
First, varying \( K_{\beta_d} \) (Figure~\ref{fig:r_vs_bd}, with \( K_{T_d}=500 \) and \( K_{\beta_s}=500 \)) shows a broadly decreasing 68\% upper limit for all three Galactic regions, with the optimum near the maximum available patches. The dust spectral index requires the finest spatial resolution.
Second, varying \( K_{T_d} \) (Figure~\ref{fig:r_vs_td}, with \( K_{\beta_d}= \)~max and \( K_{\beta_s}=500 \)) reveals that the high-latitude region (GAL020) is nearly flat with an optimum at \( K_{T_d}={300} \), the mid-latitude region requires 500 patches, and the low-latitude region requires finer resolution (optimal at 2500 patches).
Third, varying \( K_{\beta_s} \) (Figure~\ref{fig:r_vs_bs}, with \( K_{\beta_d}= \)~max and \( K_{T_d}=500 \)) produces a clear V-shape for all regions, with the optimum at low patch counts (150--300). The synchrotron spectral index varies more smoothly across the sky than the dust parameters; the \texttt{s1} model uses a $\beta_s$ template smoothed at $5^\circ$.

Combining the three regions, the optimal configuration yields a maximum-likelihood estimate of \( \hat{r} = 0.9 \times 10^{-4} \) with a 68\% confidence interval of \( [-2.7 \times 10^{-4},\; +6.5 \times 10^{-4}] \), as shown in Figure~\ref{fig:r_likelihood_optimal}.
The corresponding residual $B$-mode power spectra are shown in Figure~\ref{fig:bb_spectra_optimal}.

\begin{table}
    \centering
    \begin{tabular}{c|c|c|c}
           & \shortstack{low-lat \\ (GAL060$-$GAL040)} & \shortstack{mid-lat \\ (GAL040$-$GAL020)} & \shortstack{hi-lat \\ (GAL020)} \\
           \hline
          $K_{\beta_d}$ &  9685 (max)   & 9629 (max) & 7000 \\
          $K_{T_d}$ & 2500 & 500 & 300\\
          $K_{\beta_s}$ & 150 & 150 & 300\\
    \end{tabular}
    \caption{Optimal patch counts per spectral parameter and Galactic region, selected by minimizing the 68\% upper limit on $r$. The dust spectral index consistently requires near-maximum spatial resolution, while the synchrotron spectral index is well described by a few hundred patches. The corresponding patch geometries are visualized in Fig.~\ref{fig:patch_comparison}.}
    \label{tab:optimal_clusters_results}
\end{table}

\begin{figure*}
    \centering
    \begin{subfigure}[b]{\linewidth}
        \centering
        \includegraphics[width=0.9\linewidth]{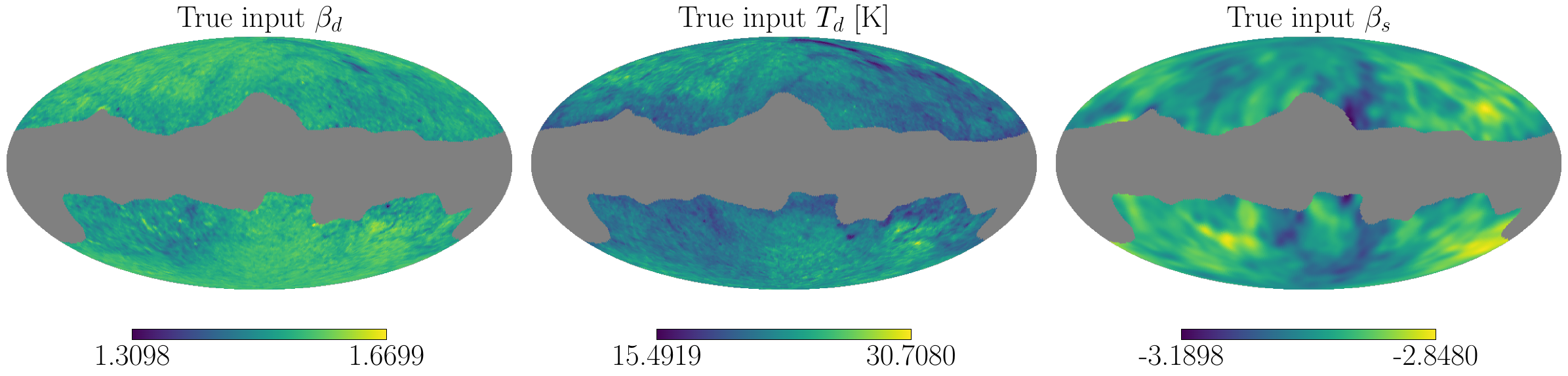}
        \caption{True input spectral parameters (\texttt{d1s1}).}
    \end{subfigure}
    \vfill
    \begin{subfigure}[b]{\linewidth}
        \centering
        \includegraphics[width=0.9\linewidth]{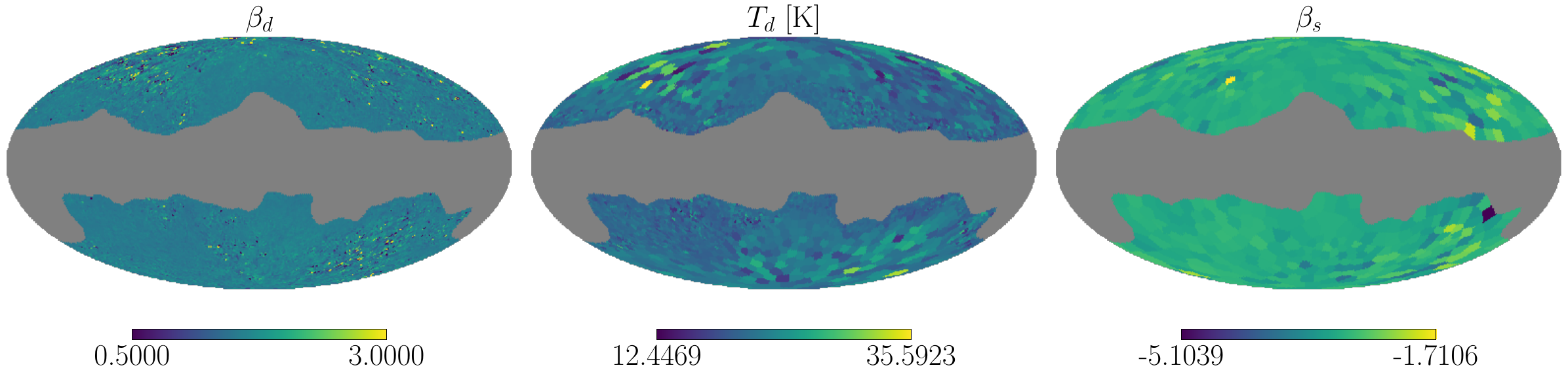}
        \caption{Recovered spectral parameters for the optimal configuration (Table~\ref{tab:optimal_clusters_results}).}
    \end{subfigure}
    \vfill
    \caption{Comparison of true input spectral parameter maps (\texttt{d1s1}, top) with recovered parameters from the optimal patch configuration (bottom), from a specific noise realisation. From left to right: dust spectral index $\beta_d$, dust temperature $T_d$, and synchrotron spectral index $\beta_s$. Grey regions are masked. The colorbar ranges differ between the input and recovered maps to accommodate the broader range of recovered values.}
    \label{fig:param_maps_comparison}
\end{figure*}

The optimal configuration recovers some of the main spatial features of the true \texttt{d1s1} spectral parameters. Figure~\ref{fig:param_maps_comparison} compares the input parameter maps with the recovered maps for the configuration of Table~\ref{tab:optimal_clusters_results} from one signal realisation, showing that some of the large-scale spatial structure of \( \beta_d \), \( T_d \), and \( \beta_s \) is qualitatively reproduced, and with significant differences between the two sets of maps due to the noise in the reconstructed one.

\subsection{K-Means Clustering vs. Multi-Resolution Superpixels}
\label{subsec:kmeans_vs_multires}

We compare our optimized K-means approach with the fixed multi-resolution configuration of~\cite{LITEBIRD_PTEP_2022}. In the latter, each spectral parameter is assigned to a predetermined \textsc{HEALPix} $N_{\rm side}$ level within each Galactic region. That configuration was chosen for its performance through manual tuning informed by inspection of the systematic residuals, not through an automated search with a selection metric applicable to real data. Our approach replaces manual selection with a systematic, metric-driven grid search.
Table~\ref{tab:multires_vs_kmeans} summarizes both setups. This comparison measures the combined benefit of (i) exhaustive optimization over $>$13{,}000 patch-count configurations and (ii) the K-means geometry, relative to a single fixed multi-resolution configuration. The improvement should not be attributed to geometry alone. Additionally, the~\cite{LITEBIRD_PTEP_2022} pipeline applies post-component separation processing steps (e.g., masking additional sky area in the reconstructed CMB map, marginalizing the reconstructed dust template in the likelihood on $r$); our comparison uses only the raw component separation output for both methods.

\begin{table}
    \centering
    \begin{tabular}{c|c|c|c}
        \hline
        & \shortstack{low-lat \\ (GAL060$-$GAL040)} & \shortstack{mid-lat \\ (GAL040$-$GAL020)} & \shortstack{hi-lat \\ (GAL020)} \\
        \hline
        $\beta_d$ & 64 (9{,}685) & 64 (9{,}629) & 64 (9{,}695) \\
        $T_d$     &  8 (255)     &  4 (88)      &  0 (1)       \\
        $\beta_s$ &  4 (88)      &  2 (34)      &  2 (23)      \\
        \hline
    \end{tabular}
    \caption{Multi-resolution configuration from~\protect\cite{LITEBIRD_PTEP_2022}, expressed in \textsc{HEALPix} $N_{\rm side}$ levels per Galactic region, with the resulting patch count in parentheses. $N_{\rm side} = 0$ denotes a uniform (single-patch) model; at $N_{\rm side} = 64$ the parameter is fit per pixel. Our optimized K-means patch counts are given in Table~\ref{tab:optimal_clusters_results}. The corresponding patch geometries are compared in Fig.~\ref{fig:patch_comparison}. Patches that intersect the mask boundary contain fewer valid pixels than the nominal $(N_{\rm side}/N_{\rm side}^{\rm target})^2$; incomplete patch counts per region are: low-lat $\beta_d{:}\ 0$, $T_d{:}\ 208$, $\beta_s{:}\ 84$; mid-lat $\beta_d{:}\ 0$, $T_d{:}\ 83$, $\beta_s{:}\ 34$; hi-lat $\beta_d{:}\ 0$, $T_d{:}\ 0$, $\beta_s{:}\ 23$.}
    \label{tab:multires_vs_kmeans}
\end{table}

\begin{figure*}
    \centering
    \begin{minipage}[t]{0.48\textwidth}
        \centering
        \includegraphics[width=\linewidth]{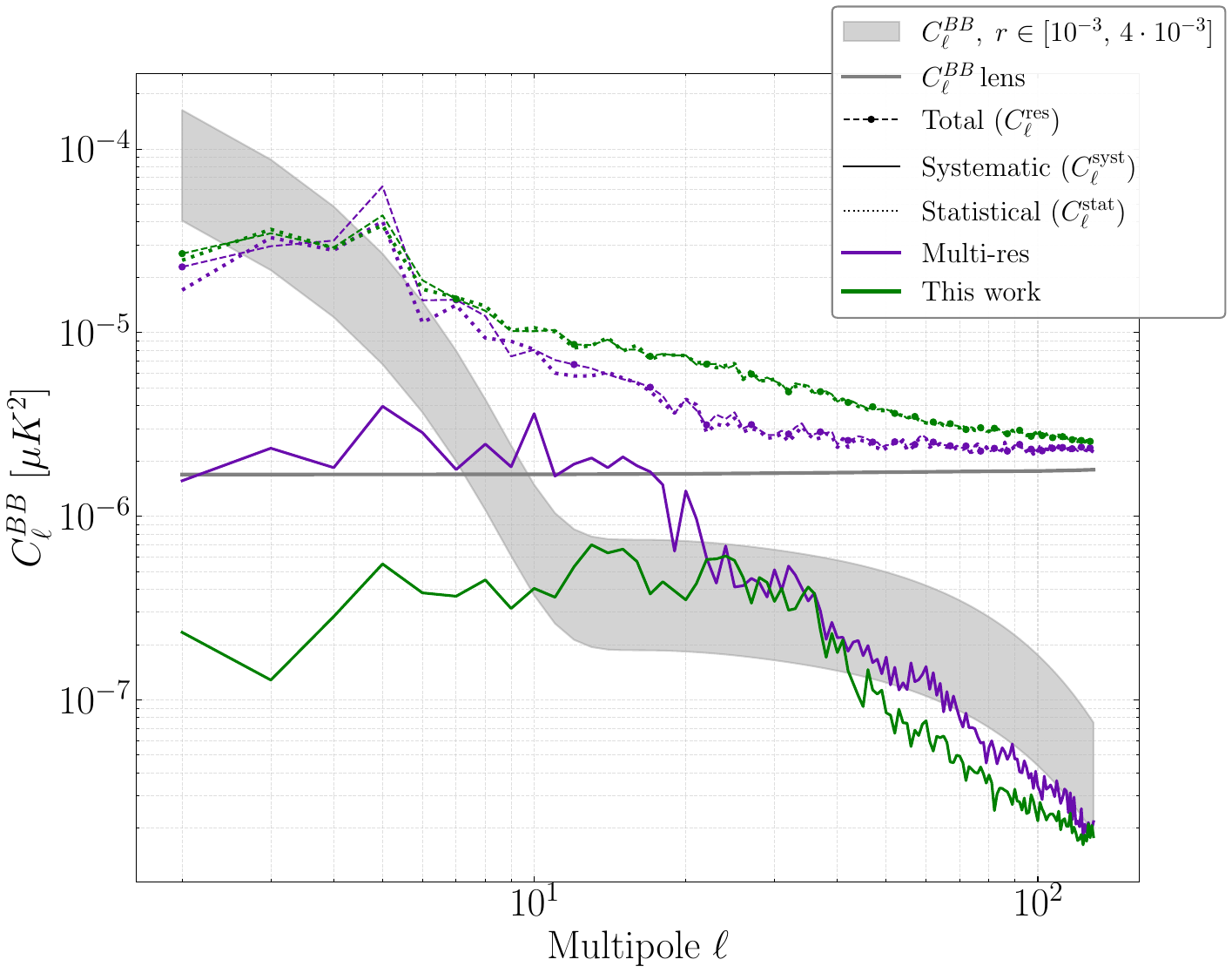}
        \caption{Residual $B$-mode power spectra comparing the multi-resolution configuration of~\protect\cite{LITEBIRD_PTEP_2022} (purple) with our optimized K-means configuration (``This work'', green), corresponding to the ``All Combined'' configuration of Fig.~\ref{fig:r_vs_clusters_seq}d. Line styles indicate total ($C_\ell^{\rm res}$, dashed), systematic ($C_\ell^{\rm syst}$, solid), and statistical ($C_\ell^{\rm stat}$, dotted) residuals. The gray band shows the primordial $C_\ell^{BB}$ for $r \in [10^{-3},\, 4 \times 10^{-3}]$; the solid gray line is the lensing contribution.}
        \label{fig:bb_spectra_comparison}
    \end{minipage}\hfill
    \begin{minipage}[t]{0.48\textwidth}
        \centering
        \includegraphics[width=\linewidth]{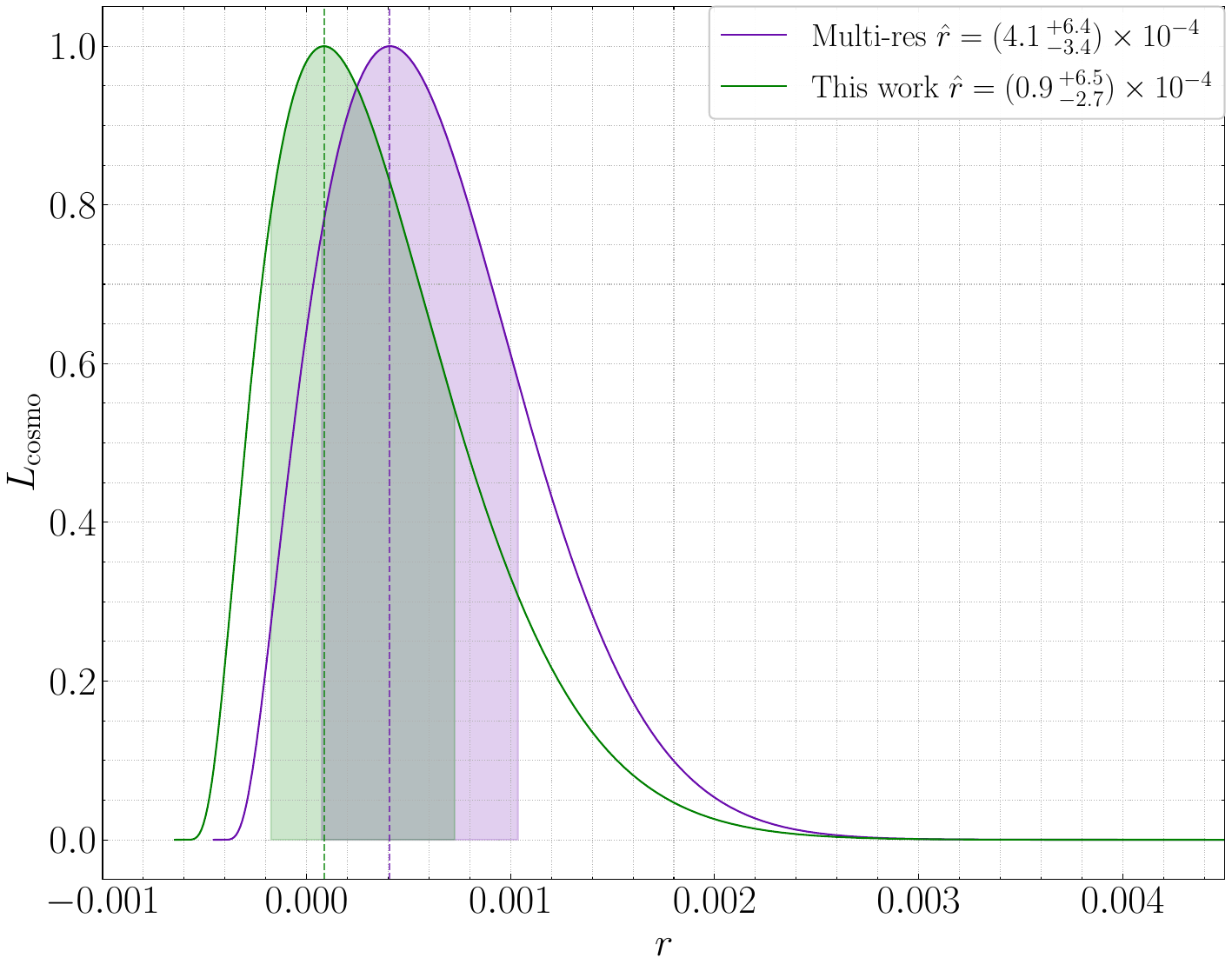}
        \caption{Posterior likelihood on $r$ for the multi-resolution configuration of~\protect\cite{LITEBIRD_PTEP_2022} (purple) and our optimized K-means configuration (``This work'', green), corresponding to the ``All Combined'' configuration of Fig.~\ref{fig:r_vs_clusters_seq}d, for an input sky with $r = 0$ (dashed black line). Dashed colored lines and shaded bands indicate the maximum-likelihood $\hat{r}$ and 68\% confidence interval for each method.}
        \label{fig:r_likelihood_comparison}
    \end{minipage}

\end{figure*}

At the power spectrum level, Fig.~\ref{fig:bb_spectra_comparison} shows that the systematic residuals (solid lines) of our K-means configuration (green) lie roughly one order of magnitude below those of the multi-resolution approach (purple) at $\ell \lesssim 20$, around the recombination bump where the primordial $B$-mode signal peaks. 
The optimized configuration's systematic residuals fall below the primordial $C_\ell^{BB}$ for $r = 10^{-3}$ for most modes.
The multi-resolution configuration's residuals overlap with or exceed $C_\ell^{BB}$ for $r\,\in\, [10^{-3},\, 4 \times 10^{-3}]$ across many multipoles of interest.
Statistical residuals (dotted lines) are slightly higher in our optimized configuration, as expected given the higher number of patches.
The total residuals (dashed line) are dominated by statistical residuals.
In summary, the K-means configuration trades systematic residuals for statistical residuals relative to the multi-resolution configuration.

\begin{figure*}
    \centering
    \begin{subfigure}[b]{0.9\textwidth}
        \centering
        \includegraphics[width=\linewidth]{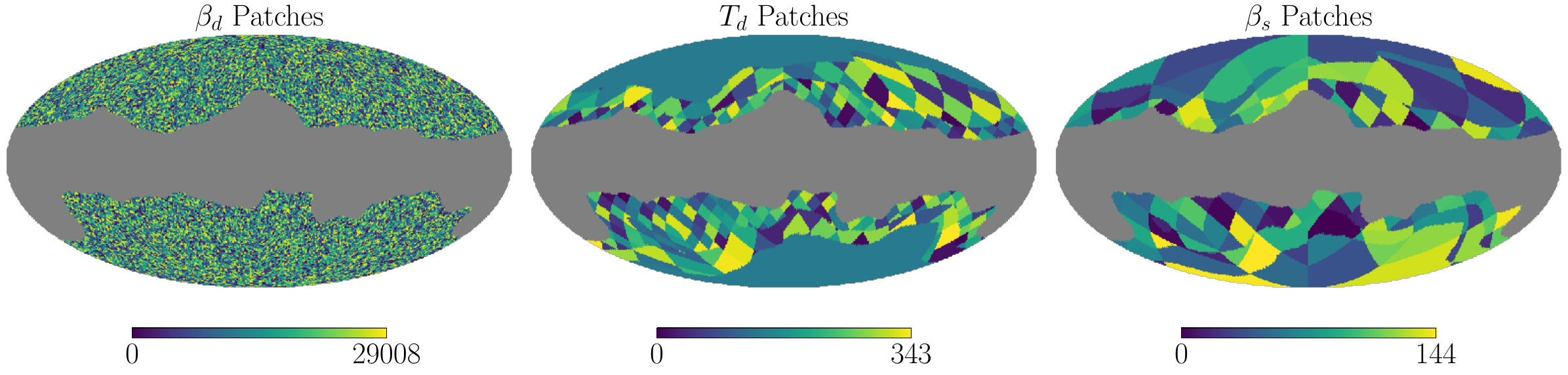}
        \caption{Multi-resolution patch assignments from~\protect\cite{LITEBIRD_PTEP_2022}: $\beta_d$ at pixel level (29{,}009 patches), $T_d$ at $N_{\rm side} \leq 8$ (344 patches), $\beta_s$ at $N_{\rm side} \leq 4$ (145 patches).}
        \label{fig:patches_multires}
    \end{subfigure}
    \hfill
    \begin{subfigure}[b]{0.9\textwidth}
        \centering
        \includegraphics[width=\linewidth]{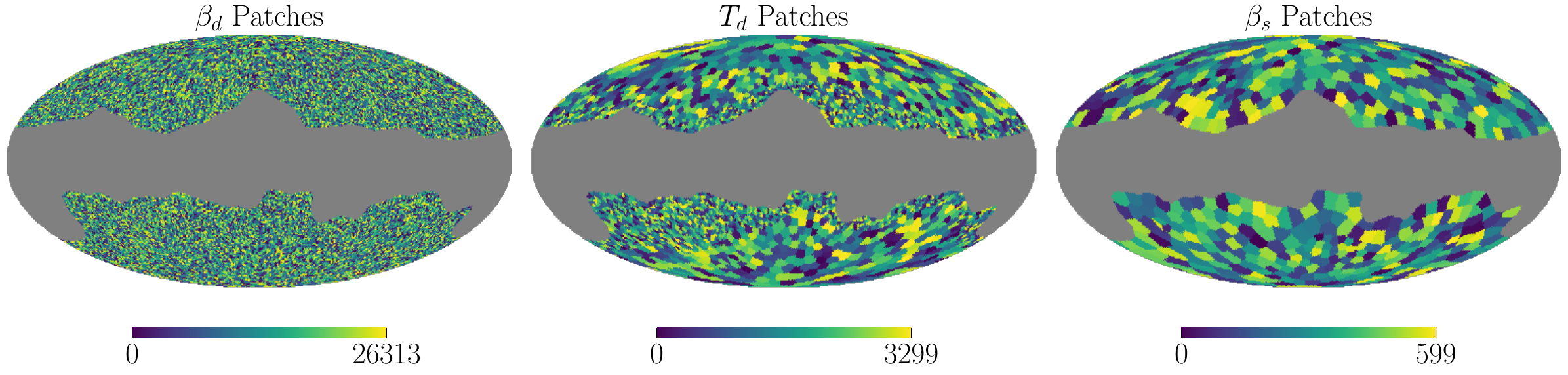}
        \caption{Optimized K-means patch assignments (this work). This map corresponds to the best fit from our grid search (Table~\ref{tab:optimal_clusters_results}), with $\beta_d$ at 26{,}314 patches, $T_d$ at 3{,}500 patches, and $\beta_s$ at 600 patches.}
        \label{fig:patches_kmeans}
    \end{subfigure}
    \caption{Comparison of patch geometries between the multi-resolution strategy of~\protect\cite{LITEBIRD_PTEP_2022} and our optimized K-means clustering, summed over the three Galactic regions. From left to right: dust spectral index $\beta_d$, dust temperature $T_d$, and synchrotron spectral index $\beta_s$. Grey regions are masked. The K-means approach allocates significantly more patches to $T_d$ and $\beta_s$, which is associated with the reduction in systematic bias. The patch counts for each method are given in Tables~\ref{tab:optimal_clusters_results} and~\ref{tab:multires_vs_kmeans}.}
    \label{fig:patch_comparison}
\end{figure*}

The cosmological likelihood level determines performance.
Figure~\ref{fig:r_likelihood_comparison} gives the $r$ constraints for both configurations. The multi-resolution approach yields $\hat{r} = 4.1 \times 10^{-4}$, indicating significant residual bias; our K-means method gives $\hat{r} = 0.9 \times 10^{-4}$, consistent with the input $r = 0$ at $0.1\sigma$. The 68\% upper limits are $\hat{r} + \sigma(r) = 1.0 \times 10^{-3}$ and $7.4 \times 10^{-4}$. Figure~\ref{fig:patch_comparison} compares the patch geometries.

Optimizing the patch configuration achieves a ${\sim}30\%$ reduction in the 68\% upper limit on $r$ compared with the fixed, manually tuned multi-resolution configuration of~\cite{LITEBIRD_PTEP_2022} for an instrument like LiteBIRD. Statistical uncertainties $\sigma(r)$ are comparable; the improvement is driven by the reduction of systematic bias. 

\subsection{Application to Non-Zero Tensor-to-Scalar Ratio}
\label{ssec:application_nonzero_r}

We validate our method on a sky model with uniform foreground scaling laws (\texttt{d0s0}) and a non-zero tensor-to-scalar ratio, \( r = 3 \times 10^{-3} \). This sky model tests whether we lose CMB signal when selecting a specific patch configuration.
Two configurations are tested: a \emph{uniform configuration} with $K_{\beta_d}=1$, $K_{T_d}=1$, $K_{\beta_s}=1$, applied to the entire sky without region partitioning (labeled ``low patches''), and an \emph{over-parameterized configuration} with $K_{\beta_d}=30{,}000$, $K_{T_d}=1{,}500$, $K_{\beta_s}=1{,}500$ (labeled ``high patches''). 
The CMB signal is preserved by Eq.~\ref{eq:CMB_unit_response} regardless of $r$ or clustering choice, so the bias on \( r \) remains near zero. We show that \( r \) is not underestimated for non-zero input $r$ and any clustering choice, justifying minimization of the $68\%$ upper limit on $r$.
Equation~\ref{eq:CMB_unit_response} holds in practice when multiplicative systematic effects are absent or correctly modelled, i.e., properly merged in the mixing matrix $\mathbf{A}$~\citep{Verges2021, Jost2023, EMA}. 

\begin{figure*}
    \centering
    
    \begin{minipage}[t]{0.45\textwidth}
        \centering
        \includegraphics[width=\linewidth]{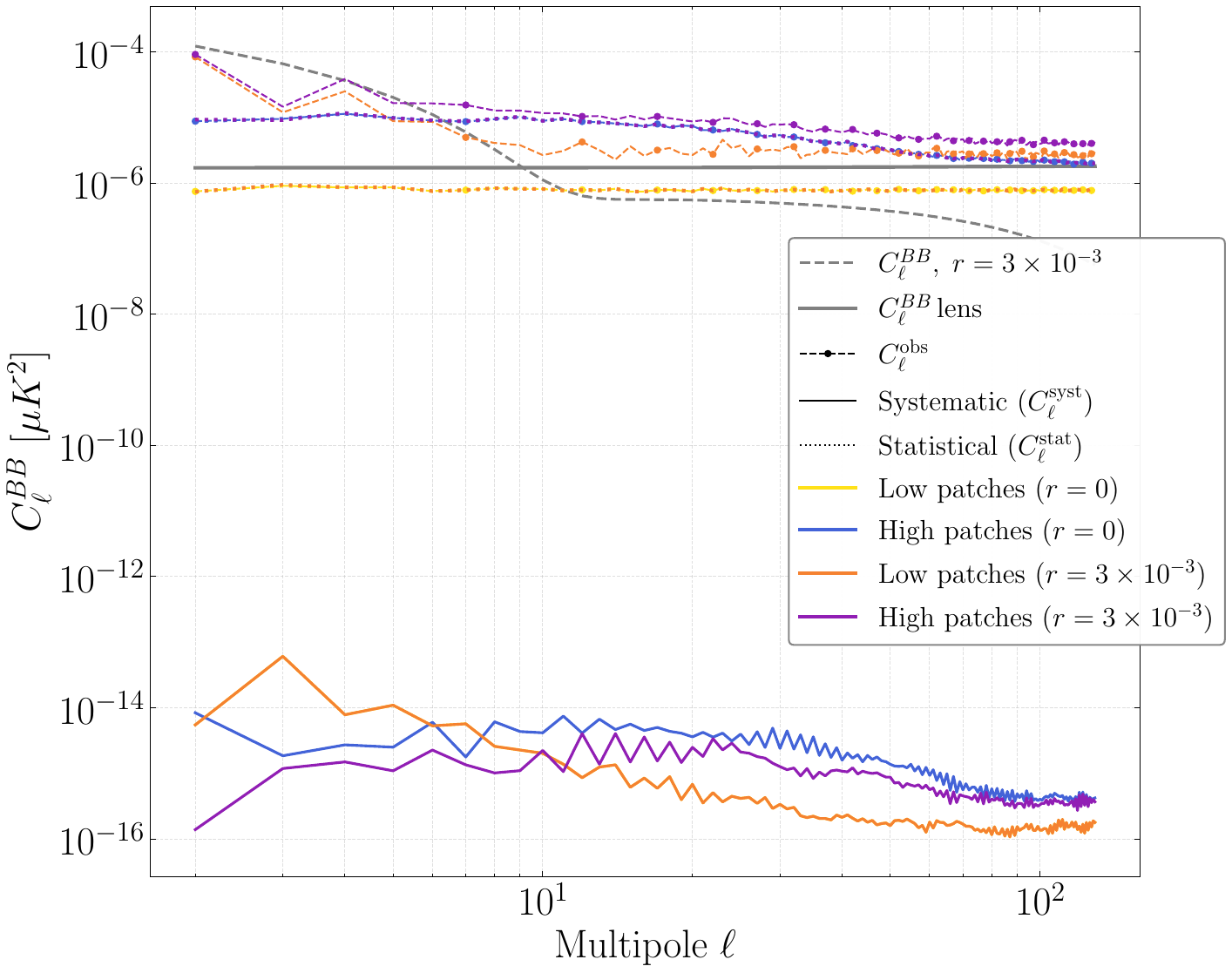}
        \caption{Residual $B$-mode power spectra for the uniform configuration (``low patches'', $K_{\beta_d}=K_{T_d}=K_{\beta_s}=1$) and an over-parameterized configuration (``high patches''), applied to input skies with $r = 0$ and $r = 3 \times 10^{-3}$. Line styles indicate observed spectra ($C_\ell^{\rm obs}$, dashed), systematic ($C_\ell^{\rm syst}$, solid), and statistical ($C_\ell^{\rm stat}$, dotted) residuals. The uniform configuration is drawn in yellow on the $r = 0$ sky and in orange on the $r = 3 \times 10^{-3}$ sky, and the over-parameterized configuration in blue and purple, respectively. The dashed grey line shows the primordial $C_\ell^{BB}$ for $r = 3 \times 10^{-3}$; the solid grey line is the lensing contribution.}
        \label{fig:bb_spectra_nonzero_r}
    \end{minipage}\hfill
    \begin{minipage}[t]{0.51\textwidth}
        \centering
        \includegraphics[width=\linewidth]{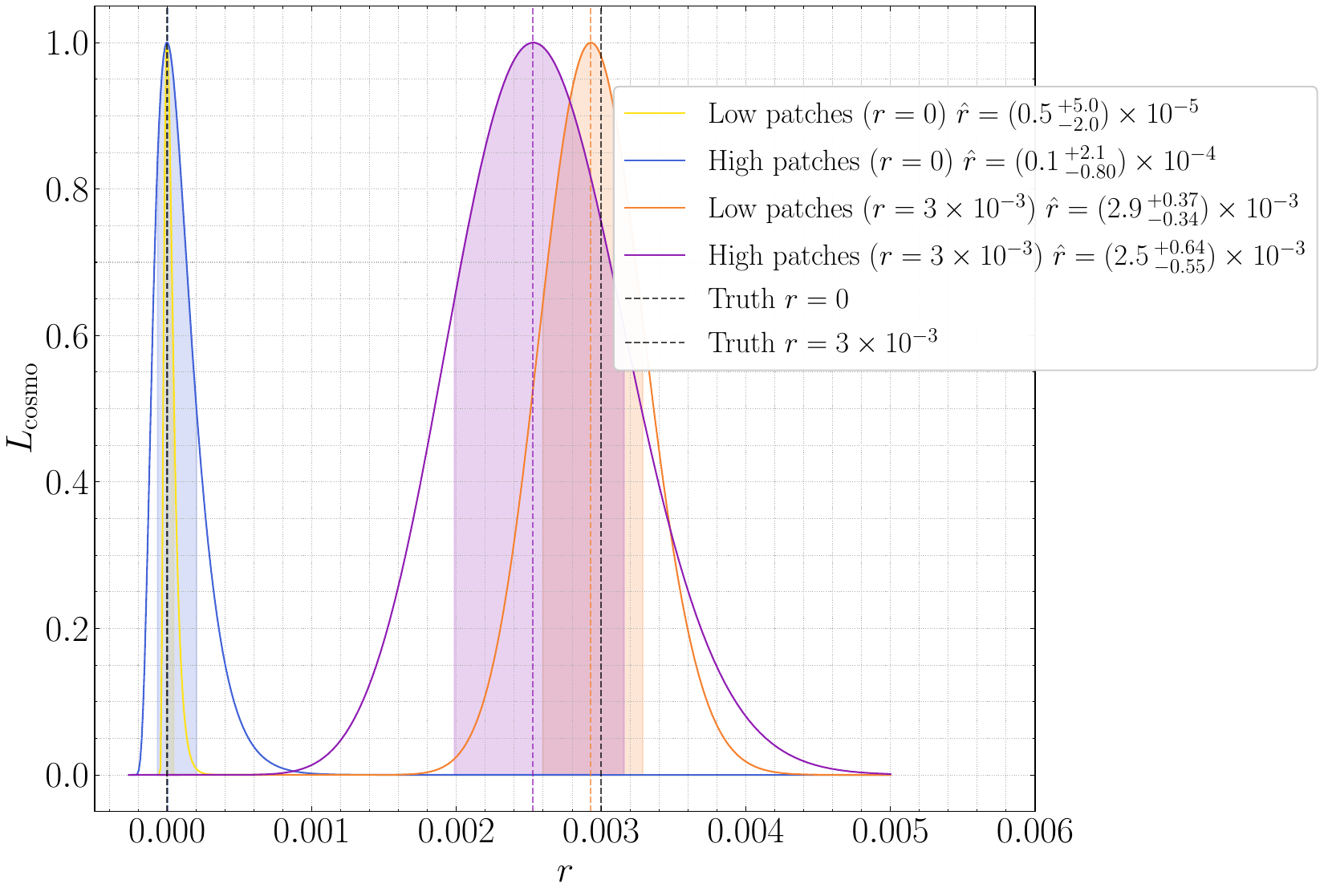}
        \caption{Posterior likelihood on $r$ for the uniform and over-parameterized configurations, applied to input skies with $r = 0$ and $r = 3 \times 10^{-3}$. Dashed colored lines and shaded bands indicate the maximum-likelihood $\hat{r}$ and 68\% confidence interval for each case, and the dashed black lines mark the input values. Colors follow Fig.~\ref{fig:bb_spectra_nonzero_r}. Both configurations recover the input $r$ within their respective 68\% intervals; the over-parameterized configuration exhibits wider uncertainties due to the increased number of degrees of freedom.}
        \label{fig:r_likelihood_nonzero_r}
    \end{minipage}

\end{figure*}

Figures~\ref{fig:bb_spectra_nonzero_r} and \ref{fig:r_likelihood_nonzero_r} show the results.
The uniform configuration applied to the \( r = 3 \times 10^{-3} \) sky (orange) exhibits total residuals comparable to its \( r = 0 \) counterpart (yellow), confirming that the non-zero primordial signal passes through the separation unaffected. 
The over-parameterized configuration gives similar results.
Fig.~\ref{fig:r_likelihood_nonzero_r} shows that the uniform configuration recovers $\hat{r} = 2.9 \times 10^{-3}$, consistent with the input $r = 3 \times 10^{-3}$ at $0.2\sigma$, with tight uncertainties (${}^{+0.37}_{-0.34} \times 10^{-3}$). The over-parameterized configuration yields $\hat{r} = 2.5 \times 10^{-3}$ with wider uncertainties (${}^{+0.64}_{-0.55} \times 10^{-3}$).
In both cases the input \( r \) remains within the 68\% confidence interval.
Over-parameterizing the foreground scaling laws recovers an unbiased estimate of \( r \), but the uniform configuration achieves the tightest constraint.

\subsection{Towards More Flexible Pixel Subsets}
\label{ssec:post_clustering}

The K-means-based grid search works for recovering an optimized configuration for future space-based missions, assuming a \texttt{d1s1} foreground sky and LiteBIRD-like bands and noise levels. This may not hold for different instrumental setups or more complex foreground models, raising the question of further improvements.

The optimized configuration exhibits statistical residuals driven by the large number of patches. Residual levels are orders of magnitude higher than the contribution from instrumental noise alone, propagated through component separation, which is of order $10^{-6} \mu K^2$.

\begin{figure}
    \centering
    \includegraphics[width=\columnwidth]{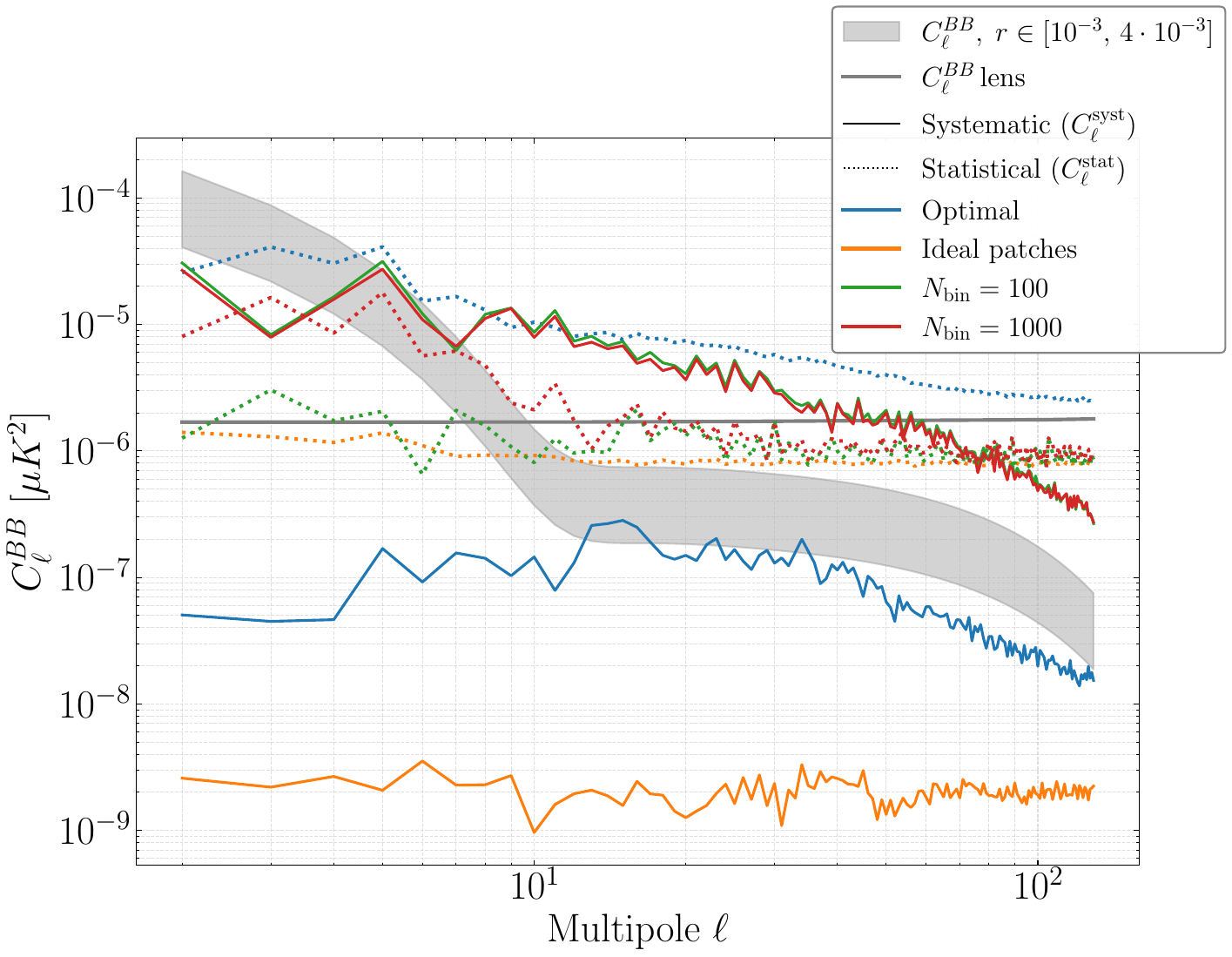}
    \caption{Statistical (dotted) and systematic (solid) $B$-mode residual power spectra for the optimal K-means configuration and three progressively coarser configurations obtained by binning the recovered spectral-parameter templates into $N_{\rm bin} = 1000$ and $100$ equal-width intervals per parameter with the statistical residuals computed by averaging across 40 noise realisations using equation~\ref{eq:stat_residuals}. As disconnected clusters are merged, statistical residuals decrease while systematic residuals increase. The gray band shows the primordial $C_\ell^{BB}$ for $r \in [10^{-3},\, 4 \times 10^{-3}]$; the solid gray line is the lensing contribution.}
    \label{fig:cls_putting_together_clusters}
\end{figure}

\cite{Rizzieri_2025} demonstrated the existence of spatially disconnected pixel subsets that perform well in terms of both statistical and systematic residuals using state-of-the-art \texttt{PySM} foreground models; these appear as ``ideal'' in Fig.~\ref{fig:cls_putting_together_clusters}.
Such configurations were identified by exploiting prior knowledge of the foreground models, specifically by binning the spectral parameter templates used to generate the foregrounds.
The open question is how to construct such subsets feasibly for real data.

\begin{figure}
    \centering
    \includegraphics[width=\columnwidth]{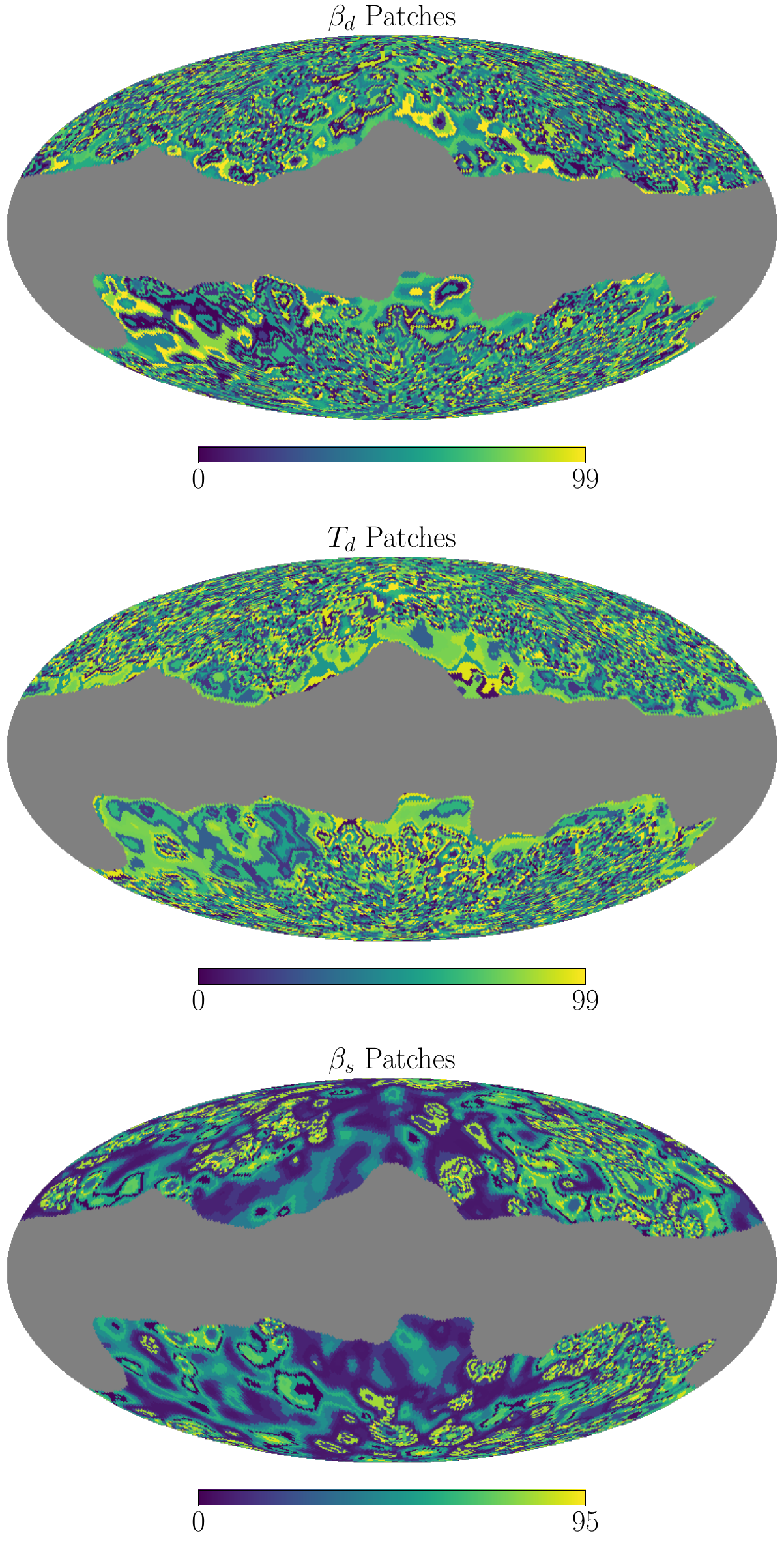}
    \caption{Sky partition obtained by binning the recovered spectral-parameter templates of the optimal K-means configuration into $N_{\rm bin}=100$ equal-width intervals per parameter ($\beta_d$, $T_d$, $\beta_s$). Pixels sharing the same bin for a given parameter are assigned to the same pixel subset, yielding spatially disconnected regions. This is the coarsest configuration shown in Figure~\ref{fig:cls_putting_together_clusters}.}
    \label{fig:patches_bin100}
\end{figure}

We explore this direction, building on our optimized configuration. A naive approach to generalizing to discontinuous pixel subsets groups multiple K-means clusters from the optimized configuration into common subsets, analogous to the ``ideal'' pixel subsets.
We bin the recovered spectral parameter templates, smoothing with \texttt{healpy.smoothing} (\texttt{fwhm} $5.0^\circ$), into equal-width intervals, aggregating clusters with similar recovered spectral parameter values (applied independently per parameter). This reduces free parameters and statistical noise.
We perform component separation on these coarser, spatially disconnected pixel subsets.
Figure~\ref{fig:patches_bin100} shows the resulting sky partition for the coarsest binning ($N_{\rm bin}= 100$), where each spectral-parameter map is reduced to at most $100$ distinct pixel subsets.
Fig.~\ref{fig:cls_putting_together_clusters} shows the recovered power spectra of statistical and systematic residuals for $N_{\rm bin}=100$.
This procedure does not perform well compared to the optimized K-means configuration. While it reduces statistical residuals, systematic residuals increase significantly. Varying the number of bins does not mitigate this; systematic residuals remain elevated.

The poor performance likely stems from noisy recovered spectral parameter templates and large K-means patches in some regions.
Figure~\ref{fig:patches_bin100} shows pixel subsets dominated by noise rather than reflecting input spectral parameter morphologies.

Exploring discontinuous pixel subsets may reduce statistical residuals without increasing systematic residuals. A more rigorous approach than the naive method here is required to incorporate this degree of freedom.
For example, we could explore configurations assembled from patches of the optimized configuration, selecting those that minimize the figure of merit on $r$.
This would avoid relying on binning noisy reconstructed spectral parameters.
Alternatively, more flexible algorithms than K-means could be employed in the initial grid search, directly incorporating the degrees of freedom associated with discontinuous pixel subsets and allowing varying areas and shapes.
Both approaches increase computational complexity and will be investigated in future work.

\section{Conclusions}
\label{sec:conclusion}
This work implements parametric component separation in the \texttt{FURAX} framework~\citep{FURAX}, a \texttt{JAX}-powered environment for CMB data analysis. The AdaTopK optimizer, an active-set, adaptive-scale algorithm for high-dimensional, bounded, noisy problems, achieves up to ${\sim}100\times$ speed-ups over the \texttt{scipy}~\citep{scipy} TNC optimizer on CPU (reflecting algorithmic improvements and CPU-to-GPU parallelism) while improving parameter recovery. We replace fixed \textsc{HEALPix} multi-resolution patches with spherical K-means clustering~\citep{Lloyd1982, DhillonModha2001}, enabling flexible, data-driven spatial partitioning of foreground spectral parameters.

The \texttt{FURAX} engine makes large-scale grid searches practical: 24 candidate values for each of three spectral parameters across three Galactic regions. We select configurations minimizing the 68\% confidence-level upper bound on $r$, balancing systematic bias against statistical uncertainty. The CMB unit response (Eq.~\ref{eq:CMB_unit_response}) preserves the CMB signal irrespective of $r$ or clustering, so optimization on $r = 0$ skies loses no generality.

Optimal patch counts (Table~\ref{tab:optimal_clusters_results}) show that $\beta_d$ requires near-pixel-level resolution (${\sim}7000$--$9685$ patches), $T_d$ is intermediate and latitude-dependent (300--2500 patches), and $\beta_s$ is smooth (${\sim}150$--$300$ patches). For $r = 0$, our configuration yields $\hat{r} = 0.9 \times 10^{-4}$ with a 68\% confidence interval of $[-2.7 \times 10^{-4},\; +6.5 \times 10^{-4}]$, a ${\sim}30\%$ reduction in the $68\%$ upper limit on $r$ versus the fixed multi-resolution configuration of~\cite{LITEBIRD_PTEP_2022} ($7.4 \times 10^{-4}$ versus $1.0 \times 10^{-3}$), before further processing. Validation with $r = 3 \times 10^{-3}$ returns $\hat{r} = 2.9 \times 10^{-3}$, consistent at $0.2\sigma$. This improvement combines exhaustive optimization over $>$13{,}000 configurations and the K-means geometry. The full grid search required ${\sim}$10 hours on 64 NVIDIA H100 GPUs (${\sim}$640 GPU-hours).

The improvement is driven by reduced systematic bias; statistical uncertainties $\sigma(r)$ are comparable. This highlights spatial variability of foreground SEDs as a limiting factor in parametric component separation. Our method achieves a lower upper limit from the parametric step alone, with implications for future CMB satellite missions such as LiteBIRD~\citep{LiteBIRD2022}. 

Future work directions:
First, make the pipeline more rigorous for real data: adopt a pseudo-$C_\ell$ estimator such as \textsc{NaMaster}~\citep{Alonso2019} for partial-sky treatment and $E$-to-$B$ purification; explore robust partial-sky likelihood approximations; scale to $N_{\text{side}} = 128$--$512$ to match upcoming satellite resolution.
Second, extend to complex foreground models (frequency decorrelation, multi-component dust) and incorporate instrumental systematics (beam asymmetries, gain drift, correlated noise) into the forward model. Mitigating such systematics is critical for real-data application.
Third, generalize the grid search to more degrees of freedom: a) Galactic mask definition; b) alternative spatial partitioning (hierarchical clustering, Support Vector Machine clustering~\citep{SVM}) for non-connected pixel subsets; c) optimize over competing SED functional forms per region.
These developments would provide a unified model-selection framework for more precise and robust $r$ estimates.

\section*{Data Availability}

The code used to generate the results presented in this paper is publicly available on GitHub at \url{https://github.com/CMBSciPol/furax-cs}. The simulation data can be reproduced using the scripts provided in the repository. Specific data products derived in this study are available from the corresponding author upon reasonable request. An interactive dashboard to explore the grid search results and visualising the component separation outputs is available online.\footnote{\url{https://askabalan-furax-cs-results.hf.space/}}

\section*{Acknowledgments}
The authors would like to thank Tran Hoang Viet, Cl\'{e}ment Leloup, Radek Stompor and Amalia Villarrubia-Aguilar for useful exchanges during the development of this work.

This work was supported by the \textsc{SciPol} project (\href{https://scipol.in2p3.fr}{scipol.in2p3.fr}), funded by the European Research Council (ERC) under the European Union’s Horizon 2020 research and innovation program (Principal Investigator: Josquin Errard, Grant agreement No. 101044073). Some of the authors also benefited from the European Union’s Horizon 2020 research and innovation program under grant agreement no. 101007633 CMB-Inflate. 

This work has also received funding by the European Union’s Horizon 2020 research and innovation program under grant agreement no. 101007633 \emph{CMB-Inflate}.

The computations were performed using HPC resources provided by the \textit{Jean Zay} supercomputer at IDRIS under allocations 2024-AD010414161R2 and 2025-A0190416919 granted by GENCI.

\FloatBarrier

\bibliographystyle{rasti}
\bibliography{furax_comp_sep}

\appendix

\section{The AdaTopK Algorithm}
\label{app:adatopk}

This appendix details the implementation of the AdaTopK optimizer, designed to handle the bounded, non-convex spectral likelihood landscape using a Top-K active set strategy.

\subsection{Internal Parameter Space}

Following the strategy of TNC~\citep{TNC}, we map the physical parameters $\mathbf{x}$ (bounded by $\mathbf{l}, \mathbf{u}$) to a normalized internal representation $\mathbf{y}$ via an affine transformation. This maps the bounds to $[0, 1]$, normalizing the optimization landscape and ensuring consistent step sizes:
\begin{equation}
    \mathbf{y} = \frac{\mathbf{x} - \mathbf{o}}{\mathbf{s}},
\end{equation}
where $\mathbf{s} = \mathbf{u} - \mathbf{l}$ is the scale vector and $\mathbf{o} = \mathbf{l}$ is the offset vector (representing the lower bound).

\subsection{Top-K Release Strategy}

At each iteration, we compute a ``release score'' for every active constraint. The score quantifies the alignment of the negative gradient $-\mathbf{g}$ with the feasible direction into the valid parameter space. For a parameter $i$ currently at a bound:
\begin{equation}
    \text{score}_i = p_i \times (-g_{\text{int}, i}),
\end{equation}
where $\mathbf{g}_{\text{int}}$ is the gradient in the internal parameter space, and $p_i \in \{-1, 0, 1\}$ is the pivot vector. $p_i = -1$ indicates the parameter is at the lower bound, $p_i = 1$ at the upper bound, and $p_i = 0$ indicates it is free.

We then use JAX's hardware-accelerated \texttt{top\_k} primitive to identify the $K$ parameters with the highest positive scores (i.e., those most strongly pushed towards the feasible region by the gradient) and release them simultaneously ($\mathbf{p}_i \leftarrow 0$).

\subsection{Impact of the Top-K Fraction}
\label{app:topk_impact}

The Top-K fraction $K$ controls the number of bound constraints released per iteration and thus the subspace exploration rate. We evaluated $K$ from 0.0 to 1.0 on the \textit{Jean Zay} HPC cluster (NVIDIA H100 GPUs), with a model comprising ${\sim}\,30{,}000$ $\beta_d$ parameters and $1{,}500$ parameters each for $\beta_s$ and $T_d$, totaling ${\sim}\,33{,}000$ free parameters.

At $K{=}1.0$ (all constraints released simultaneously), the rapid changes in the active set induce ``chattering''~\citep{Kelley_1999}; parameters oscillate between active and inactive states across iterations, destabilizing convergence and, in several cases, leading to worse local minima. Higher $K$ values can reduce per-iteration cost by releasing constraints in batches, but the resulting convergence behaviour is erratic: nearby $K$ values sometimes lead to substantially different final objective values, making the optimizer difficult to tune reliably.

At $K{=}0$ (single-parameter release per iteration, matching the classical TNC strategy), convergence is the most predictable and consistently reaches the lowest objective values. Although individual iterations release fewer constraints, the stability gain more than compensates in practice.

Based on these results, we adopt $K{=}0$ for all production runs reported in this paper. The Top-K machinery remains available as a configurable hyperparameter for future exploration on problems where faster active-set turnover may be beneficial.

\subsection{Dynamic State Rescaling}

To prevent numerical underflow in low-SNR regions (high latitudes) or overflow in high-SNR regions (Galactic plane), we monitor the gradient norm $\|\mathbf{g}\|$. If it falls outside a safe numerical range (typically $[10^{-15}, 10^{15}]$), we rescale the cost function by a factor $f_{\text{scale}} = 1/\|\mathbf{g}\|$. 

To preserve the optimizer's trajectory, we also rescale the internal moment estimates ($\mathbf{m}, \mathbf{v}$) of the underlying AdaBelief solver:
\begin{equation}
    \mathbf{m} \leftarrow f_{\text{scale}} \cdot \mathbf{m}, \quad \mathbf{v} \leftarrow f_{\text{scale}}^2 \cdot \mathbf{v}.
\end{equation}
This ensures the optimizer continues seamlessly across the vast dynamic range of the sky signal without resetting its momentum.

\subsection{Algorithm Summary}

\begin{algorithm}
\caption{AdaTopK Optimization Step}
\label{alg:adatopk}
\begin{algorithmic}[1]
\State \textbf{Input:} Params $\mathbf{x}$, Bounds $\mathbf{l}, \mathbf{u}$, Gradients $\mathbf{g}$
\State \textbf{Init:} Compute scale $\mathbf{s}$, offset $\mathbf{o}$, pivot $\mathbf{p}$
\State Map to internal space: $\mathbf{y} \leftarrow (\mathbf{x} - \mathbf{o}) / \mathbf{s}$, $\mathbf{g}_{\text{int}} \leftarrow \mathbf{g} \cdot \mathbf{s}$
\State \textbf{Release Constraints:}
    \State \quad Compute scores: $S_i \leftarrow p_i \cdot (-g_{\text{int}, i})$
    \State \quad Identify Top-$K$ indices where $S_i > 0$
    \State \quad Set $p_i \leftarrow 0$ for Top-$K$ indices
\State \textbf{Project Gradients:} $\mathbf{g}_{\text{proj}} \leftarrow \mathbf{g}_{\text{int}} \odot (\mathbf{p} == 0)$
\State \textbf{Rescale State:}
    \If{$\|\mathbf{g}_{\text{proj}}\|$ outside safe range}
        \State $f \leftarrow 1/\|\mathbf{g}_{\text{proj}}\|$
        \State Scale $\mathcal{L}, \mathbf{g}_{\text{proj}}, \mathbf{m}, \mathbf{v}$ by $f, f, f, f^2$
    \EndIf
\State \textbf{Compute Direction:}
    \State \quad $\mathbf{d}, (\mathbf{m}, \mathbf{v}) \leftarrow \text{AdaBelief}(\mathbf{g}_{\text{proj}}, \text{state})$
    \State \quad Mask direction: $\mathbf{d} \leftarrow \mathbf{d} \odot (\mathbf{p} == 0)$
\State \textbf{Step Limit:} $\alpha_{\max} \leftarrow \text{DistanceToNearestBound}(\mathbf{y}, \mathbf{d})$
\State \textbf{Line Search:} Find $\alpha \in [0, \alpha_{\max}]$ minimizing $\mathcal{L}$
\State \textbf{Update:}
    \If{$\alpha = \alpha_{\max}$}
        \State $\mathbf{y} \leftarrow \mathbf{y} + \alpha \mathbf{d}$ (Clamp to bound)
        \State Update $\mathbf{p}$ for limiting parameter
    \Else
        \State $\mathbf{y} \leftarrow \mathbf{y} + \alpha \mathbf{d}$
    \EndIf
\State \textbf{Output:} Physical params $\mathbf{x} \leftarrow \mathbf{y} \cdot \mathbf{s} + \mathbf{o}$
\end{algorithmic}
\end{algorithm}

\section{The \texttt{furax-cs} Package}
\label{app:package}

The \texttt{furax-cs} package is an open-source Python implementation of the framework described in this paper. It is available on the Python Package Index (\texttt{pip install furax-cs}) and built on top of JAX \citep{JAX} and FURAX, inheriting full GPU acceleration and JIT compilation. The source code is hosted on GitHub.\footnote{\url{https://github.com/CMBSciPol/furax-cs}}

\subsection{Package Structure}

The package is organised into three core modules:

\begin{description}
  \item[\texttt{furax\_cs.data}] Simulate and cache multi-frequency sky maps using PySM3 and CAMB, load instrument specifications, and handle Galactic masks.
  \item[\texttt{furax\_cs.optim}] Unified \texttt{minimize()} interface supporting multiple solvers (L-BFGS, AdaTopK, SciPy back-ends), box constraints via sigmoid reparameterisation, and full \texttt{vmap} compatibility for parallel optimisation across grid points.
  \item[\texttt{furax\_cs.kmeans\_clusters} / \texttt{multires\_clusters}] Adaptive K-means clustering and multi-resolution partitioning of the sky into spectrally homogeneous regions.
\end{description}

An additional \texttt{r\_analysis} subpackage provides post-component separation utilities for pseudo-$C_\ell$ estimation of the tensor-to-scalar ratio $r$ and publication-ready plotting.

\subsection{Usage Example}

The three-step pipeline (clustering, noise generation, optimisation) is illustrated below:

\begin{lstlisting}
# Step 1: Adaptive K-means clustering
clusters = kmeans_clusters(key, mask, indices, patches)

# Step 2: Instrumental noise operator
noised_d, N, _ = generate_noise_operator(
    key, noise_ratio, indices, nside,
    masked_d, instrument, "QU")

# Step 3: Bounded optimisation (AdaTopK, Appendix A)
params, state = minimize(
    fn=negative_log_likelihood_fn,
    init_params=init_params,
    solver_name="ADABK0",
    lower_bound=lower, upper_bound=upper,
    nu=nu, N=N, d=noised_d,
    patch_indices=clusters)

# Recover component signals
S = sky_signal_fn(params, nu=nu, d=noised_d,
                  N=N, patch_indices=clusters)
\end{lstlisting}

The \texttt{minimize} call returns the optimised spectral parameters for each sky patch. The \texttt{sky\_signal} function from FURAX then uses these parameters to solve the mixing system and recover the component signals $\hat{\mathbf{s}}$, including the CMB.

\subsection{Command-Line Interface}

For batch execution on HPC clusters, \texttt{furax-cs} exposes several CLI entry points: \texttt{kmeans-model}, \texttt{ptep-model}, \texttt{fgbuster-model}, and \texttt{r\_analysis}, among others. These wrap the Python API with SLURM-compatible argument parsing for large-scale parameter sweeps. Full documentation is available in the repository.

\label{lastpage}
\end{document}